\documentclass[]{interact}
\usepackage[caption=false]{subfig}
\usepackage[T1]{fontenc}
\usepackage{lmodern}
\usepackage{listings}
\usepackage{enumitem}
\setlist[itemize]{itemsep=0.5\baselineskip, topsep=0.5\baselineskip}
\newcommand{\rev}[1]{#1}
\renewcommand{\line}[1]{Line~\ref{line:#1}}
\newcommand{\Line}[1]{Line~\ref{line:#1}}
\newcommand{\lines}[2]{Lines~\ref{line:#1}--\ref{line:#2}}
\newcommand{\Lines}[2]{Lines~\ref{line:#1}--\ref{line:#2}}
\usepackage[natbibapa,nodoi]{apacite}
\usepackage[british]{babel}
\usepackage{microtype}

\usepackage{etoolbox}
\apptocmd{\thebibliography}{\interlinepenalty=10000\relax}{}{}

\usepackage{xcolor}
\usepackage{graphicx}
\usepackage{comment}
\usepackage{longtable}
\usepackage{silence}
\usepackage{caption}
\usepackage{array}
\usepackage{float}
\usepackage[per-mode=symbol]{siunitx}
\newcommand{\figdir}{figures}
\usepackage{lststyle}
\newcommand{\codedir}{code}
\definecolor{amber}{HTML}{FF5F00}
\let\origthelstnumber\thelstnumber
\makeatletter
\newcommand*\nonum{%
  \lst@AddToHook{OnNewLine}{%
    \let\thelstnumber\relax%
     \advance\c@lstnumber-\@ne\relax%
    }%
}
\renewcommand*\num{
  \lst@AddToHook{OnNewLine}{%
   \let\thelstnumber\origthelstnumber%
   \advance\c@lstnumber\@ne\relax}%
}
\newcommand*\numdecr{\num\global\advance\c@lstnumber-\@ne\relax}
\makeatother
\usepackage{fancyhdr}
\fancypagestyle{conflictpage}{
  \fancyhf{}
  \fancyhead[L]{Conflict of Interest: The authors declare no conflict of interest.}
  
}
\usepackage{cleveref}

\begin{document}
\title{A Practical Guide to PID Controller Implementation\thanks{This work was partially supported by the Wallenberg AI, Autonomous Systems and Software Program (WASP) funded by the Knut and Alice Wallenberg Foundation. All authors from Lund University are members of the ELLIIT Strategic Research Area. Moreover, the paper was also partially funded by the Spanish Ministry of Science within the project PID2023-150739OB-I00.}}
\author{\name{E. Sundström\textsuperscript{a}, M. Bauer\textsuperscript{b}, J. L. Guzm\'an\textsuperscript{c}, T. H\"agglund\textsuperscript{a} and K. Soltesz\textsuperscript{a}}\affil{\textsuperscript{a}Dept. of Automatic Control, Lund University, Box 118, SE-22100 Lund, Sweden \\(email: \{emil.sundstrom, tore.hagglund, kristian.soltesz\}@control.lth.se)\\ \textsuperscript{b}Dept. of Process Engineering, Hamburg University of Applied Sciences, Hamburg, Germany (email: margret.bauer@haw-hamburg.de)\\ \textsuperscript{c}Dept. of Informatics, Universidad de Almer\'ia, ceiA3, CIESOL, Ctra. Sacramento s/n, 04120 Almer\'ia, Spain (email: joseluis.guzman@ual.es)}}
\maketitle
\begin{abstract}
How difficult can it be to implement a PID controller? The answer is twofold. Implementing the PID control law is simple and computationally inexpensive. However, this basic form will not work in practical applications. The primary reason for this is the various physical limitations of the actuator. Measurement noise, different implementations depending on the various structures (P, PI, PD or PID), bumpless transfer, and varying sampling interval also result in problems rendering the basic form inoperable. PID implementation is therefore more difficult than meets the eye. This paper introduces a reference implementation of the PID controller which considers these practical issues. It includes pseudo-code, discussion of the implementation choices and simulation of carefully selected, important test cases.
\end{abstract}
\section{Introduction}\label{sec:intro}
The future of PID control was correctly predicted in an article 25 years ago \citep{astrom+2001}: it is still the most prevalent control strategy in real world applications. PID is a superior strategy when assessed in terms of performance, tuning, ease of use and maintenance, resulting in a good solution for many process dynamics.
Then and now, the PID implementation has been buried in proprietary archives of control technology suppliers. From MathWorks' MATLAB implementations to automotive, flight control, process control systems, etc.: the PID controller is generally implemented as a black box, and the inside workings are not disclosed to the user. In some instances, the archives are not well maintained and the PID controller is poorly implemented with serious problems. The knowledge of how to implement the PID control law correctly is neither communicated nor developed.
\clearpage
The classical PID control law in the form
\begin{equation}
u(t)=K\left(e(t)+\frac{1}{T_i}\,\int_0^t e(\tau)\, d\tau+T_d\frac{de(t)}{dt}\right)\label{eq:pidTimeDomain}
\end{equation}
is easy to implement in hardware and software \citep{hagglund+2024}. \rev{While the nomenclature of \cref{eq:pidTimeDomain} is well-established, specifications of involved signals and parameters are listed in \cref{sec:nomenclature}, alongside those of other nomenclature introduced throughout the paper.}
While it is possible to implement the PID algorithm in analogue electronic circuits, mechanics, pneumatics and even biological systems, the most prevalent implementation is in programming code. The ease of implementation may explain why there appears to be only one publication solely dedicated to PID implementation \citep{clarke1984}, covering some basic steps focusing on the $z$-domain. Several textbooks cover some basic aspects, too, e.g., \cite{shinskey1994,shinskey1996,astrom+2006,visioli2006}.
In recent years, the void of PID implementation publications has been filled with a large number of internet educational resources providing implementations of the PID controller in various languages (C, C++, Python or MATLAB). Unfortunately, these implementation resources realise \cref{eq:pidTimeDomain} but rarely include the practical considerations required when implementing a real-world PID controller.
Some publications describe student projects or experimental setups that implement the PID controller, see \citep{bhandari+2022,uzunovic+2010,krejcar+2011}. A clear description of how to implement the real-world PID controller is not available.
Some efforts have been made to standardise the PID controller. The IEC has published a standard for evaluating the performance of PID control \citep{iec2010}.
The International Society of Automation (ISA) has issued a technical report on PID algorithms and performance that standardises PID control, focusing on nomenclature \citep{isa2023}. There is currently no standard covering PID controller implementation.
These circumstances make it difficult for control engineering professionals to find a concise and thoroughly explained reference implementation. As a result, PID implementations in commercial solutions are often ad hoc and may not include important features required to deal with practical problems. Anecdotally, the authors have heard of several industrial controllers without anti-windup that could not be used during process startup.
There are numerous extensions to the standard controller, including fractional-order controllers \citep{chen+2009}, PID controllers that additionally consider the second derivative of the error \citep{huba+2018}, etc. These fall outside the scope of this work. Neither do we consider PID tuning. \rev{Instead, the sole focus of this paper lies in arriving at an algorithmic structure for a digital reference implementation of the PID controller.}
There are two alternative implementation forms: positional and incremental, the latter also referred to as velocity form \citep{clarke1984}. Both are explained in detail in \cref{sec:digimp}. \Cref{sec:features} first describes the most common problems found in practice, extending the basic discretisation of the PID control law to deal with all of them effectively. The result, and our main contribution, is the reference implementation given in \cref{sec:CombinedPID}, which combines both positional and incremental form. \rev{The positional form is needed when there is no integral action, while the incremental form inherently provides bumpless behaviour when there is. Unlike the incremental form, the positional form could in principle reproduce the full functionality of the combined form, but only at the cost of special-case handling and storing numerous past signal values. The combined form provides this functionality without these drawbacks.}
\clearpage

The reference implementation is provided in pseudo-code, discussed line by line, aided by a graphical representation of the code. While this means that the code cannot be taken directly into one development environment such as MATLAB, C, or Python, the readability is improved because the code provides a clearer description of the algorithm, without any distractions.
\Cref{sec:closedloopimp} introduces additional implementation considerations such as the state initialisation, the execution interval and the runtime environment. The most common problems that were described in \cref{sec:features} are then simulated in \cref{sec:basicexamples} to show the benefit of extending the basic PID algorithm to practical working code. The simulation codes are available in a GitHub repository by \cite{sundstrom+2026} located at \url{https://github.com/copybit/PID.code}. \rev{A tutorial-oriented webpage by \cite{sundstrom+2026b} that documents and exemplifies our reference implementation is found at \url{https://w3.ual.es/personal/joguzman/pid/}.}
\section{Digital PID implementation}\label{sec:digimp}
\begin{figure}[ht]
\centering
\includegraphics[width=0.45\textwidth]{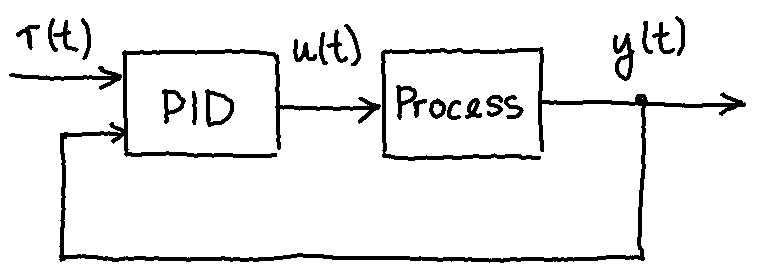}
\caption{The basic control loop.}
\label{fig:BasicLoop}
\end{figure}

To implement the controller in a computer, \cref{eq:pidTimeDomain} must be converted to a discrete time version. The principle of discrete time sampling is explained in this section. Both the integral and the derivative parts of the PID controller must be approximated for the conversion. There are two alternative implementations of the PID controller: the positional form and the incremental form \citep{clarke1984}. Both will be described in the following. The positional form is more intuitive, but the incremental form has important practical advantages. The implementation presented in this paper will be based on a combination of the two to exploit the advantages of each form.
The naming of variables and parameters in this implementation is based on the standard form described in \cref{eq:pidTimeDomain} where $u(t)$ is the control signal and $e(t)=r(t)-y(t)$ is the control error, that is, the difference between the setpoint $r(t)$ and the measured process variable $y(t)$.
The basic control loop is depicted in \cref{fig:BasicLoop}. Note that setpoint $r(t)$ and process variable $y(t)$ enter the PID block instead of control error $e(t)$. The reason for this depiction will become apparent in the following sections.
The parameters of the PID controller in \cref{eq:pidTimeDomain} are controller gain $K$, integral time $T_i$, and derivative time $T_d$. Our implementation uses the representation
\begin{equation}
u(t)=k_p e(t)+k_i\,\int_0^t e(\tau)\, d\tau+k_d\frac{de(t)}{dt}\label{eq:pidParallel}
\end{equation}
with parameters $k_p=K$, $k_i=K/T_i$, and $k_d=K T_d$, since it is more general in the sense that it allows for $T_i=0$ in \cref{eq:pidTimeDomain} without causing a division-by-zero error. If other forms are preferred, such as the series implementation \citep{astrom+2006}, the implementation codes presented in this paper can be easily adapted.
\subsection{Discrete time sampling}\label{sec:sampling}
\begin{figure}[ht]
\centering
\includegraphics[width=0.45\textwidth]{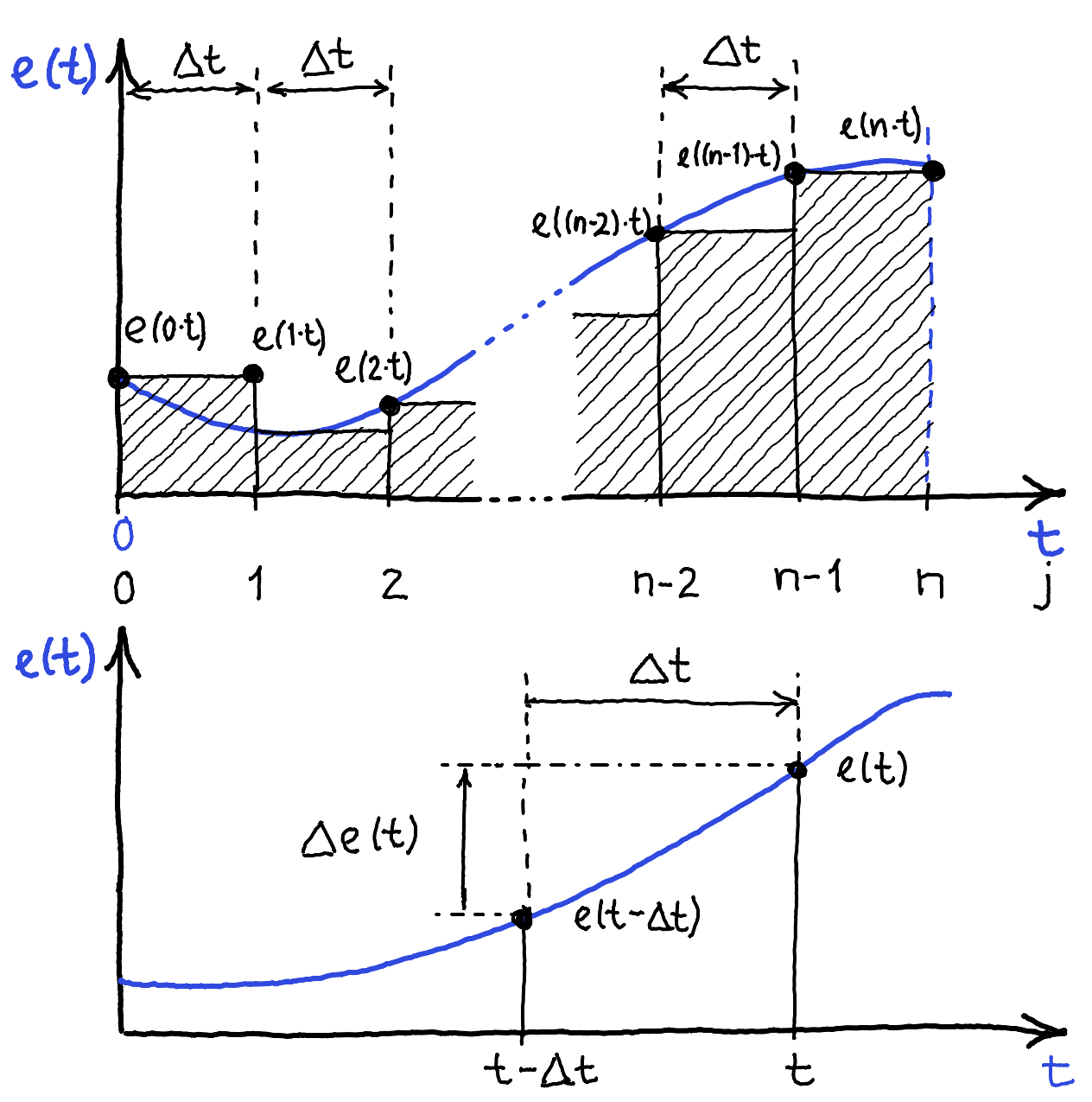}
\caption{Approximations of the integral and the derivative.}
\label{fig:Discretization}
\end{figure}

Time series sampling is the prerequisite for digital implementation of \cref{eq:pidTimeDomain}. The continuous-time process variable $y(t)$ and the continuous-time setpoint $r(t)$ have to be sampled at discrete time intervals $\Delta t$. The discrete-time control error is then
\begin{equation}
    e(t) = r(t) - y(t),
\end{equation}
with $t=n \Delta t$, where $n$ is the discrete time index. The control signal $u(t)$ is then calculated at discrete times $n\Delta t$.
\Cref{eq:pidTimeDomain} contains the integral and the derivative of the control error $e(t)$. A common way of discretising integral and derivative is to approximate the integral with a sum and the derivative with a difference such that
\begin{subequations}\label{eq:approximations}
\begin{align}
\int_{0}^{t} e(\tau)\,d\tau
&\approx \Delta t \sum_{j=0}^{n} e(j\Delta t),\\[2pt]
\frac{de(t)}{dt}
&\approx \frac{e(t)-e(t-\Delta t)}{\Delta t}
 = \frac{\Delta e(t)}{\Delta t}.
\end{align}
\end{subequations}
These approximations are illustrated in \cref{fig:Discretization}. The integral is approximated using the forward rectangular rule, and the derivative is approximated using the forward difference approximation. Note that also other approximations such as backward difference or tangent line approximation can be considered without affecting the results presented in this paper.
\subsection{Positional form}\label{sec:positional}
Introducing the approximations from \cref{eq:approximations} in \cref{eq:pidTimeDomain}, the discrete version of the PID controller results in the positional form
\begin{equation}
    u(t) = k_p\, e(t) + k_i \Delta t \sum^{n}_{j=0}e(j\Delta t) + k_d\ \frac{e(t) - e(t-\Delta t)}{\Delta t},
    \label{eq:posForm1}
\end{equation}
where proportional gain $k_p=K$, integral gain $k_i=K/T_i$, and derivative gain $k_d=KT_d$
have been introduced. Splitting \cref{eq:posForm1} into corresponding terms yields
\begin{equation}
u(t)=u_p(t)+u_i(t)+u_d(t),
\label{eq:PIDterms}
\end{equation}
where $u_p(t)$, $u_i(t)$, and $u_d(t)$ represent the proportional, integral, and derivative terms, respectively. A more efficient way to implement the PID controller is to update the integral $u_i(t)$ recursively, and rewrite the individual terms as follows
\begin{subequations}\label{eq:posForm2}
\begin{align}
u_p(t) &= k_p\, e(t),\\[8pt]
u_i(t) &= u_i(t-\Delta t)+k_i\, \Delta t\, e(t),\\[4pt]
u_d(t) &= k_d \, \dfrac{e(t)-e(t-\Delta t)}{\Delta t}
       = k_d\frac{\Delta e(t)}{\Delta t}.
\end{align}
\end{subequations}
PID controllers are often implemented based on the positional discrete representation of \cref{eq:PIDterms} and \cref{eq:posForm2}. This representation is also sometimes called the direct form. In each iteration, the control signal is computed from the two most recent setpoint and process variable values, with the previous value of the error $e(t-\Delta t)$ and the previous value of the error integral $u_i(t-\Delta t)$ as controller states.
\subsection{Incremental form}\label{sec:incremental}
The incremental form, also called the velocity form \citep{clarke1984}, computes the control signal increment since the last iteration, and uses the control signal itself to store the controller state. The control law of the incremental form is
\begin{equation}
    u(t) = u(t-\Delta t) + \Delta u(t),
    \label{eq:incrementLaw}
\end{equation}
where $\Delta u(t)$ represents the increment of the signal $u$ between time $t-\Delta t$ and time $t$:
\begin{equation}
\Delta u(t)=u(t)-u(t-\Delta t).
\end{equation}
(This use of $\Delta$ applies also to other signals than $u$.) The control signal increment $\Delta u(t)$ can be split into the individual terms
\begin{equation}
\Delta u(t) = \Delta u_p(t) + \Delta u_i(t) + \Delta u_d(t),
\label{eq:DeltaUTerms}
\end{equation}
which can be obtained from \cref{eq:posForm2} as
\begin{subequations}\label{eq:incformLong}
\begin{align}
\Delta u_p(t) &= k_p\,\Delta e(t),\\[8pt]
\Delta u_i(t) &= k_i\, \Delta t\, e(t),\\[4pt]
\Delta u_d(t) &= k_d\, \Big( \dfrac{\Delta e(t)}{\Delta t}
                  - \dfrac{\Delta e(t-\Delta t)}{\Delta t}\Big),
\end{align}
\end{subequations}
where $\Delta e(t-\Delta t)=e(t-\Delta t)-e(t-2\Delta t)$. In each iteration, the term increments are calculated according to \cref{eq:incformLong} from the three most recent error values, with the actual control signal itself, $u(t)$, saved as a controller state between iterations.

\subsection{Incremental versus positional form}\label{sec:incvcpos}
The incremental form has two important practical advantages over the positional form: bumpless transfer and clamping anti-windup are easily implemented.

Bumpless transfer---a requirement in most applications---means that switching between manual and automatic or between tracking and automatic mode, or changing controller parameters, does not cause large jumps in the controller output $u$. Positional form implementations achieve this by detecting such changes and adjusting the integrator state accordingly. In the incremental form, the output only changes by the increment $\Delta u$, so no such checks and adjustments are needed.

The integral part can give rise to integral windup: when the actuator and process cannot achieve the desired setpoint, the persistent control error may keep increasing the control signal beyond the physical limitations of the actuators. Three common remedies are control signal clamping, back calculation and integrator clamping. All can work in practice, but back calculation requires tuning an additional parameter, a tracking-time constant. In the incremental form, control signal clamping is obtained simply by keeping the control signal inside the control limits.

On the other hand, the incremental form requires the integral term and therefore cannot be used for P- or PD-controllers (\cref{sec:PandPD}); in these cases the positional form is required, handled in the implementation of \cref{sec:CombinedPID}. This is a minor drawback, since neither anti-windup nor bumpless transfer are relevant without integral action.

Pseudo-code for the two forms, based on \cref{eq:posForm2} and \cref{eq:incformLong}, is presented in \cref{code:position} and \cref{code:incremental}, respectively. We drop the time index, replacing $u(t)$ with \lstinline{u}, etc. Previous values, needed by the increments, carry the prefix \lstinline{x}; for example, $e(t-\Delta t)$ is written as \lstinline{xe}. The letter \lstinline{D} denotes $\Delta$, so $\Delta e(t) = e(t)-e(t-\Delta t)$ becomes \lstinline{De=e-xe}. The pseudo-code language is detailed in \cref{sec:language}. \rev{The purpose of \cref{code:position,code:incremental} is a simple comparison between the positional and incremental forms; several practical features, including initialisation of state variables, have therefore been omitted. These are covered in-depth for the combined form PID controller of \cref{sec:CombinedPID}.}
\noindent\makebox[\linewidth][c]{
\begin{minipage}[t]{0.44\linewidth}
\begin{lstlisting}[
  caption={Basic positional form.},
  label={code:position}
]
function u=control(r,y) |\nonum|
  % Compute control signal |\num|
  e=r-y
  up=kp*e
  ui=xui+ki*Dt*e
  ud=kd*(e-xe)/Dt
  u=up+ui+ud |\nonum|

  % Update state |\numdecr|
  xe=e
  xui=ui
end
\end{lstlisting}
\end{minipage}\hspace{0.04\linewidth}
\begin{minipage}[t]{0.51\linewidth}
\begin{lstlisting}[
  caption={Basic incremental form.},
  label={code:incremental}
]
function u=control(r,y) |\nonum|
    % Increment control signal |\num|
    e=r-y
    De=e-xe
    Dup=kp*De
    Dui=ki*Dt*e
    ud=kd*(e-xe)/Dt
    Dud=ud-xud
    u=xu+Dup+Dui+Dud |\nonum|
    
    % Update state |\numdecr|
    xe=e
    xud=ud
    xu=u
end
\end{lstlisting}
\end{minipage}
}
\section{Implementation features}\label{sec:features}
Implementing the basic form of the PID controller is easy, but practical problems complicate the basic code. These problems often only become apparent during development and are usually not taught in a course or included in textbooks. Some but not all problems may occur simultaneously. Conversely, not all adjustments may be necessary. However, there is no major downside in any of the discussed modifications. \Cref{tab:problems} gives an overview of the problems that are addressed in the reference implementation presented in \cref{sec:CombinedPID} and are discussed in the remainder of this section.
\begin{table}
    \centering
       \caption{Overview of practical implementation problems and their solutions addressed in \cref{sec:features}.}
       \renewcommand{\arraystretch}{1.5}
    \begin{tabular}{>{\raggedright\arraybackslash}p{0.54\textwidth}>{\raggedright\arraybackslash}p{0.27\textwidth}c}
    \toprule
        \textbf{Problem} &\textbf{Solution}  &\textbf{Section}  \\
        \midrule
        If the controller has \textbf{no integral} action, $k_i=0$, then the stationary control error is zero only when control signal $u=0$.
        & Add a control signal bias term $u_0$ when $k_i=0$. & \ref{sec:PandPD} \\
        \textbf{Setpoint changes} lead to large proportional and derivative control signal terms, potentially damaging the actuator. & Setpoint weights &  \ref{sec:setpoint} \\
        Adding a \textbf{feed-forward} signal to the control signal of the PID controller results in integral windup. & Add the feed-forward signal to the control signal before applying the anti-windup strategy.  & \ref{sec:FF} \\ 
        Controller signal $u$ may exhibit larger than permissible \textbf{jumps} from one time step to the next. & Rate limitation  & \ref{sec:ulims} \\
        \textbf{Integral windup}: The integral part increases, winding up the control signal beyond the actuator’s saturation limit. Once wound up, the control signal will take a long time to wind down again. The control signal will stay saturated during wind-down.  & Anti-windup & \ref{sec:windup} \\
        A controller in \textbf{tracking mode}, whose control signal is not applied, may experience integral windup. & Controller output tracking  & \ref{sec:tracking}  \\
        Control signal may jump when \textbf{switching mode} or when \textbf{changing parameter} values. & Bumpless transfer  & \ref{sec:manauto} \\
        \textbf{Execution interval $\Delta t$ may vary} and the control signal may be computed incorrectly. & Jitter compensation & \ref{sec:jitter} \\
        \textbf{High frequency noise} of the process variable and in rarer cases the setpoint can lead to undesired movements of the derivative and proportional parts of the control signal. & Process variable and setpoint filtering & \ref{sec:filter} \\
        \bottomrule
    \end{tabular}
  \label{tab:problems}
\end{table}
\subsection{Control without integral action}\label{sec:PandPD}
\begin{figure}[h]
\centering
\includegraphics[width=0.7\textwidth]{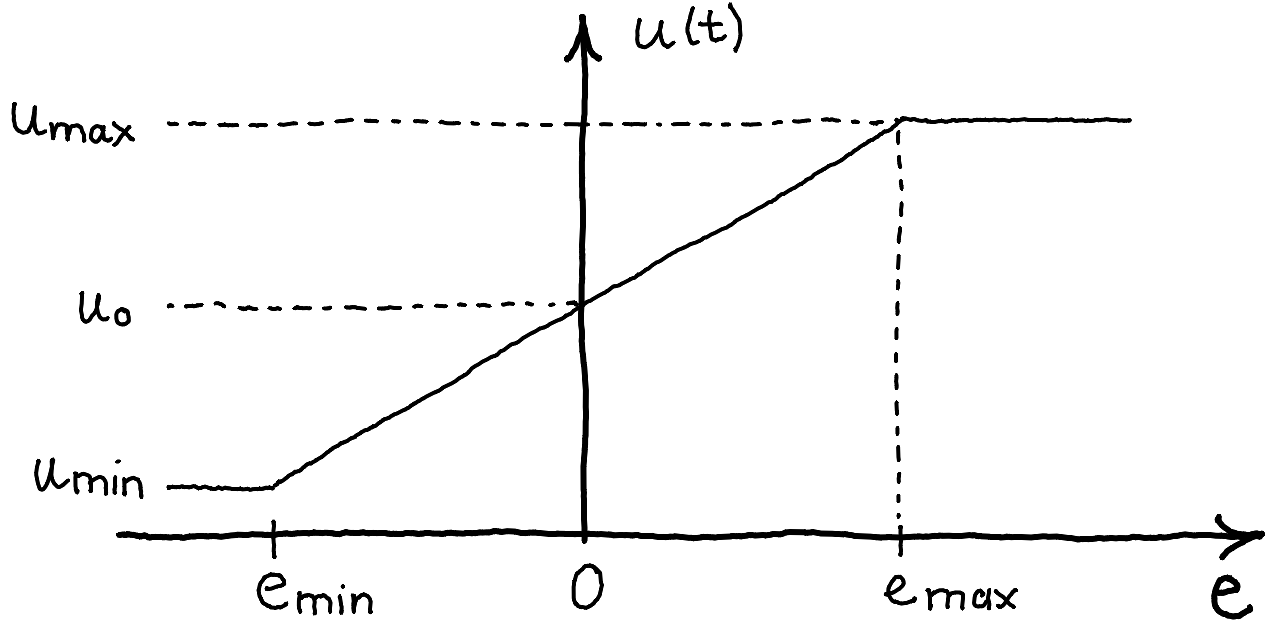}
\caption{Relationship between control error $e(t)$ and control signal $u(t)$ for a P-controller.}
\label{fig:PControl}
\end{figure}

The PID implementation must be able to handle the situation of a PID control law without integral action, such as pure P- or PD-control. As discussed earlier, the incremental form cannot be used for controllers without integral action because the incremental form results in a non-changing control signal when the error is constant. To see this phenomenon, observe \cref{eq:incrementLaw,eq:DeltaUTerms,eq:incformLong} when $k_i=0$ and the error is constant. Then $\Delta u(t)$ is zero, no matter the value of the control signal $u(t)$ and the constant error $e(t)$. We therefore risk getting ``stuck'' far from $e(t)=0$. Also, consider a parameter update of $k_p$ for a P-controller when the error is constant, but non-zero. In incremental form, the control action stays the same after the update, and the stationary error of the P-controller is unaffected, which should not be the case.
Therefore, the desired strategy for our PID implementation to handle P- and PD-control is to operate the controller in positional form whenever $k_i=0$. This is illustrated in the example given in \cref{sec:example2}. A bias term $u_0$ must be added to reduce stationary errors, see e.g. \citep{astrom+2006}. The bias term is nothing more than a constant control signal, that is normally adjusted to the control signal corresponding to zero error. This is illustrated in \cref{fig:PControl} for a P-controller, which shows how the value of $u_0$ encodes the offset from $u(t)=0$ when the error is zero. The slope in the figure corresponds to the proportional gain $k_p$, and the limits $u_{\min}$ and $u_{\max}$ represent the actuator limitations.

Thus, when there is no integral action in the controller, the integral term $u_i(t)$ is replaced by $u_0$, such that the control signal in this situation becomes
\begin{equation}
    u(t)=u_p(t)+u_0+u_d(t).
    \label{eq:posFormBias}
\end{equation}
\rev{In other words, when not in manual mode, our proposed controller has a bias term that equals the user-specified parameter $u_0$ when there is no integral action ($k_i=0$), and that equals the (implicitly defined) integral control signal term $u_i(t)$ when integral action is enabled ($k_i\neq0$). The user can change $u_0$ also when integral action is enabled, but the effect of such a change will only become visible once integral action is disabled.}
\subsection{Setpoint handling}\label{sec:setpoint}
\begin{figure}[ht]
\centering
\includegraphics[width=0.5\textwidth]{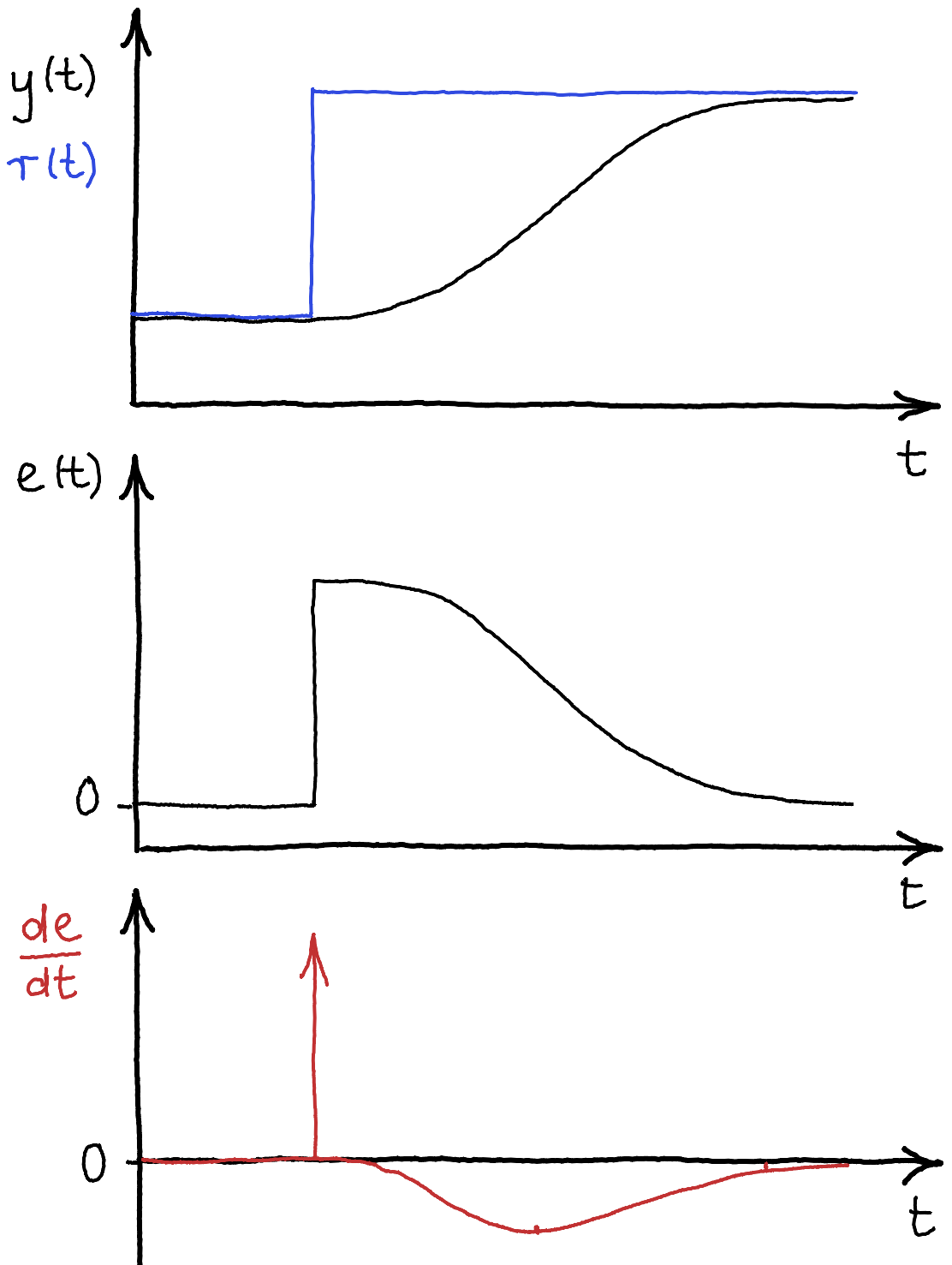}
\caption{Responses resulting from a step change in the setpoint.}
\label{fig:SetpointStep}
\end{figure}

With the positional form control laws of \cref{eq:pidTimeDomain,eq:posForm1}, a step in the setpoint $r(t)$ results in a corresponding step in the control error $e(t)=r(t)-y(t)$ and therefore a large impulse in the derivative of the control error $de(t)/dt=dr(t)/dt-dy(t)/dt$. This is illustrated in \cref{fig:SetpointStep}, from which it easily can be seen that we directly after the setpoint change get a large spike in the control signal, sometimes referred to as {\it derivative kick}.
This behaviour is normally not desired, so to obtain a smoother control signal, it is common to introduce setpoint weights in the proportional and derivative parts, so that the control error in the proportional part is replaced by $b\ r(t)-y(t)$ and the error in the derivative part by $c\ dr(t)/dt-dy(t)/dt$. The setpoint weights $b$ and $c$ are restricted to $0 \le b \le 1$ and $0 \le c \le 1$. With these modifications, \cref{eq:pidTimeDomain} becomes
\begin{equation}
    u(t) = K \left( b\ r(t)-y(t) + \frac{1}{T_i} \int_{0}^{t} e(\tau) \, d\tau + T_d \left(c\ \frac{dr(t)}{dt}-\frac{dy(t)}{dt}\right)\right),
    \label{eq:pidTimeDomain2}
\end{equation}
and implementations \cref{eq:posForm2} and \cref{eq:incformLong} are
\begin{subequations}\label{eq:posForm2SP}
\begin{align}
u_p(t) &= k_p\, \left(br(t)-y(t)\right),\\[8pt]
u_i(t) &= u_i(t-\Delta t)+k_i\, \Delta t\, e(t),\\[4pt]
u_d(t) &= k_d \,\left(c\,\dfrac{\Delta r(t)}{\Delta t}
           - \dfrac{\Delta y(t)}{\Delta t}\right),
\end{align}
\end{subequations}
for the positional form, and
\begin{subequations}\label{eq:incformLongSP}
\begin{align}
\Delta u_p(t) &= k_p\,(b\Delta r(t)-\Delta y(t)),\\[8pt]
\Delta u_i(t) &= k_i\,\Delta t\, e(t),\\[4pt]
\Delta u_d(t) &= k_d\left(
    c\,\dfrac{\Delta r(t)-\Delta r(t-\Delta t)}{\Delta t}
    - \dfrac{\Delta y(t)-\Delta y(t-\Delta t)}{\Delta t}
\right),
\end{align}
\end{subequations}
for the incremental form.
Note also that for controllers without integral action there will normally be stationary control errors. To minimise these, the choice $b=1$ is recommended to make the P part act on the true control error. When integral action is considered, the parameter $b$ can be used to reduce the impact of sudden changes in the setpoint signal on the control action, mitigating overshoot without compromising the performance of the disturbance rejection. Use of setpoint weighting is illustrated in the example given in \cref{sec:example3}. For the PID reference implementation proposed in this article, setpoint weights are included, as well as the choice $b=1$ when no integral action is present.
\subsection{Feed-forward control}\label{sec:FF}
\begin{figure}[ht]
\centering
\includegraphics[width=0.35\textwidth]{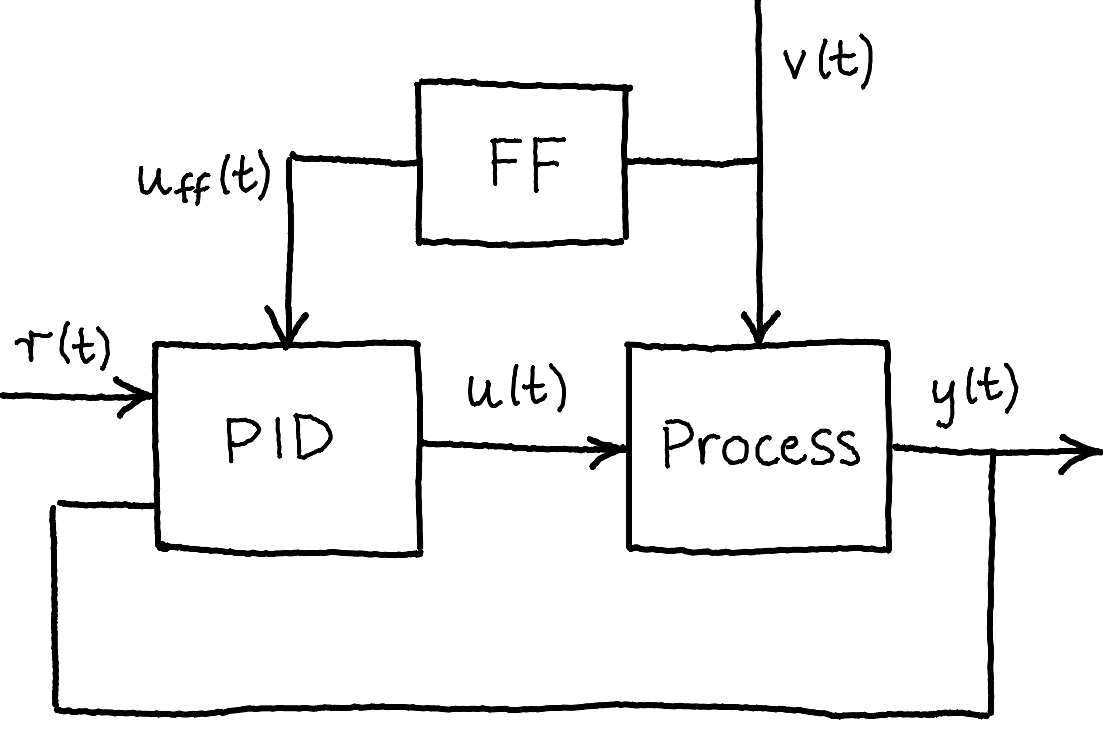}
\caption{PID controller with feed-forward from load disturbance $v(t)$.}
\label{fig:FeedForward}
\end{figure}

Feed-forward control is a complementary control strategy to the feedback strategy used in PID-control to compensate measurable load disturbances, as illustrated in \cref{fig:FeedForward}. For several reasons including anti-windup treatment, the feed-forward control signal should be added to the feedback control signal inside the PID controller, since it should affect anti-windup in the same way as other control signal terms; see \citep{guzman+2024}.
The feed-forward input signal can be used for other purposes than load disturbance compensation. It can for instance be used for improved setpoint handling and for decoupling in MIMO systems (see \citealp{liu+2019}). For a PID implementation in incremental form, the desired behaviour is that feed-forward enters with the increments
\begin{equation}
\Delta u_{ff}(t)=u_{ff}(t)-u_{ff}(t-\Delta t),
\label{eq:Duff}
\end{equation}
such that the total update equation for the control signal becomes
\begin{equation}
\Delta u(t) = \Delta u_p(t) + \Delta u_i(t) + \Delta u_d(t) + \Delta u_{ff}(t).
\end{equation}
In scenarios in which the positional form is used, the feed-forward signal then enters as a direct signal, and not through increments. Feed-forward control is illustrated in the example given in \cref{sec:example4}.
\subsection{Control signal limitation}\label{sec:ulims}
In almost all PID control systems, the controller output is limited due to actuator constraints. In many cases, the lower bound is $u_{\min}=0$~\si{\percent}, which reflects a motor being turned off or a valve being fully closed, and the upper bound is $u_{\max}=100$~\si{\percent}, which reflects the motor running at maximum speed or the valve being fully open. The values of these limits depend on the actuator for each particular control problem, and the limitation is expressed as
\begin{equation}
    u_{\min}\leq u(t) \leq u_{\max}.
    \label{eq:ulimits}
\end{equation}
In some PID control systems, the rate at which the control signal is allowed to change may also be limited due to safety or due to \rev{a slow actuator}. This puts the restriction
\begin{equation}
    \dot{u}_{\min}\leq \dot{u}(t) \leq \dot{u}_{\max},
    \label{eq:ratelim}
\end{equation}
which is normally referred to as rate limitation.
The amplitude and rate constraints can be combined at every sampling instant to form an admissible interval for the control signal. The rate limitation implies that the control signal at the current instant must satisfy
\begin{equation}
u(t-\Delta t) + \Delta t\,\dot{u}_{\min} \le u(t) \le
u(t-\Delta t) + \Delta t\,\dot{u}_{\max}.
\end{equation}
This interval is intersected with the actuator saturation limits. The resulting
admissible bounds used by the controller are therefore
\begin{equation}
u_{s,\min} = \max\!\left(u_{\min},\, u(t-\Delta t) + \Delta t\,\dot{u}_{\min}\right),
\label{eq:usmin}
\end{equation}
\begin{equation}
u_{s,\max} = \min\!\left(u_{\max},\, u(t-\Delta t) + \Delta t\,\dot{u}_{\max}\right).
\label{eq:usmax}
\end{equation}
The control signal is finally constrained to satisfy
\begin{equation}
    u_s(t) =
    \left\{
    \begin{aligned}
        &u_{s,\max} \quad \quad \quad \, \, \text{if} \quad u(t) > u_{s,\max}, \\
        &u_{s,\min} \quad \quad \quad \, \, \, \text{if} \quad u(t) < u_{s,\min}, \\
        & u(t) \quad \quad \quad\quad \,\,\text{else}.
    \end{aligned}
    \right.
\label{eq:uIncSat}
\end{equation}
This ensures that the control signal simultaneously respects the
actuator saturation limits and the maximum allowed rate of change between
consecutive sampling instants.
Saturation handling is illustrated in the examples given in \cref{sec:example2} and \cref{sec:example4}.
\subsection{Integrator anti-windup}\label{sec:windup}
Enforcing limits on the control signal means that the saturated control signal $u_s(t)$ may differ from the calculated nominal control signal $u(t)$. \rev{If a control error remains, the magnitude of the integral state will keep increasing---winding up---without resulting in any change of the already saturated control signal.} This gives rise to the well-understood phenomenon of integrator wind-up, as explained in, e.g., \citet{astrom+2006}. There are several established strategies to address integrator wind-up \rev{\citep{bohn+1995,dasilva+2018,hoyo+2023,visioli2003,skogestad2023}}, but since naming of anti-windup strategies is somewhat ambiguous in the literature, we list here the terminology used throughout this article of the most commonly used anti-windup strategies, and what those strategies mean for an incremental form implementation:
\begin{itemize}
    \item \textit{Control signal clamping}: If the calculated control signal falls outside the interval [$u_{s,\min}$, $u_{s,\max}$], it is adjusted to the saturated value, i.e., $u(t)=u_s(t)$.
    \item \textit{Back calculation}: The calculated control signal $u$ is dynamically adjusted such that the saturation error $u(t)-u_s(t)$ decays as a first-order process with a time constant $T_t$, whenever the control signal is saturated. The time constant $T_t$ is typically set to the integral time $T_i$.
    \item \textit{Integrator clamping}: In its simplest form, integrator clamping entails not updating the integrating part whenever there is saturation. In incremental form, this corresponds to setting $\Delta u_i(t)=0$ whenever the control signal saturates.
\end{itemize}
As stated in \cref{sec:incvcpos}, control signal clamping results naturally with the incremental form by saturation of the control signal according to \cref{eq:uIncSat}. However, we propose back calculation for this article due to its flexibility. In this context, flexibility means that with $T_t=\Delta t$ we have control signal clamping, but by adjusting $T_t$, different anti-windup behaviours can be obtained. As discussed in \cref{sec:ulims}, enforcing the saturation limits achieves both absolute control signal limits and rate limitations.
\rev{A slightly more advanced form of integrator clamping allows for integration exactly up to a saturation limit instead of setting the increment $\Delta u_i=0$ if it would nominally pass the limit. This is covered by the two first points in the slightly more advanced combined anti-windup strategy specified below:}
\begin{itemize}
    \item \textit{Integrate to saturation}:
    Consider the situation where the control signal $u(t)$ is saturated, but not the control signal without $\Delta u_i(t)$. Then the desired behaviour is to integrate to the limit. Thus, $\Delta u_i(t)$ is reduced.
    \item \textit{Allow integration in the right direction}: When $u(t)>u_{\max}$ then integration further into saturation ($\Delta u_i(t) > 0$) is not allowed, but integration away from saturation ($\Delta u_i(t) < 0$) is. In this case, $\Delta u_i(t)$ is not set to zero, but remains as calculated in integration. Of course, the corresponding logic applies to the lower limit $u_{\min}$.
    \item \textit{Back calculation}: The saturation error is updated according to the description of \textit{Back calculation} above.
\end{itemize}
\rev{The two first points can be used alone to achieve integrator clamping that occurs exactly at the saturation limit. They can also be part of a combined anti-windup strategy, that includes also the third point, constituted by back-calculating anti-windup. \clearpage
The rationale behind applying integrator clamping prior to back-calculation is to prohibit an initial increment integrating further into a saturated region, before the back-calculation correction is applied. By doing this, the saturation error will decay exactly following the back-calculation time constant $T_t$, whereas allowing for adding the integral increment prior to applying back-calculation may slightly shift the decay rate.}
Anti-windup is illustrated in the example given in \cref{sec:example4}.
\subsection{External signal tracking}\label{sec:tracking}
There are situations in which the controller output should track an external signal instead of trying to make the process variable follow a setpoint. Examples are override control, where another controller may take over the control, often handled by selectors, and gain scheduling using several controllers. In these instances, the controller, which currently does not affect the process, must enter \textit{tracking mode (track)} to avoid windup and bumps; see \citep{hagglund2023}. A tracking example including a simulation when tracking is desired is provided in \cref{sec:example7}.
In tracking mode, the controller assumes that the control signal currently applied to the actuator was produced by itself, even if it was not. In other words, the controller updates its internal states so that its computed output is consistent with the actual actuator signal. This means that the control signal is calculated by
\begin{equation}
    u(t)=u_{\text{track}}(t)+\Delta u_p(t) + \Delta u_i(t) + \Delta u_d(t) + \Delta u_{ff}(t),
    \label{eq:track}
\end{equation}
where $u_{\text{track}}$ is the control signal from the controller that was actually in control of the actuator in the last iteration. Controller output tracking is illustrated in the examples given in \cref{sec:example6} and \cref{sec:example7}. \rev{Note that control signal increments are added to the tracking signal $u_{\text{track}}$ in \cref{eq:track}. If this was not done, the (unsaturated) output of the tracking controller would be identical to the tracking signal $u_{\text{track}}$, preventing controller switching in selector scenarios as the one of \cref{sec:example6}.}
\vspace*{-1em}
\subsection{Bumpless mode and parameter changes}\label{sec:manauto}
There are three control modes: automatic or switched on (auto), tracking (track) as described in the previous section, and manual (man), where the control signal is specified by the user. Switching between these modes should not result in a large change or \textit{bump} in the control signal $u(t)$. Bumpless transfer is a fundamental concept in PID control and is therefore a must in a good implementation. Bumpless transfer is desired for all controller mode changes (when possible) and for controller parameter updates.
A common implementation of bumpless transfer in positional form is to adjust the integral term to obtain the desired control signal. In incremental form, since the integral state is encoded into the control signal, bumpless transfer results naturally in many situations. However, there are some scenarios where bumpless transfer is not possible without large workarounds. For example, bumpless parameter updates when $k_i=0$ are usually neither possible nor desired. \Cref{tab:bumplessReq} describes when bumpless transfer is desired for the PID reference implementation and when not.

\begin{figure}[H]
\centering
\includegraphics[width=0.7\textwidth]{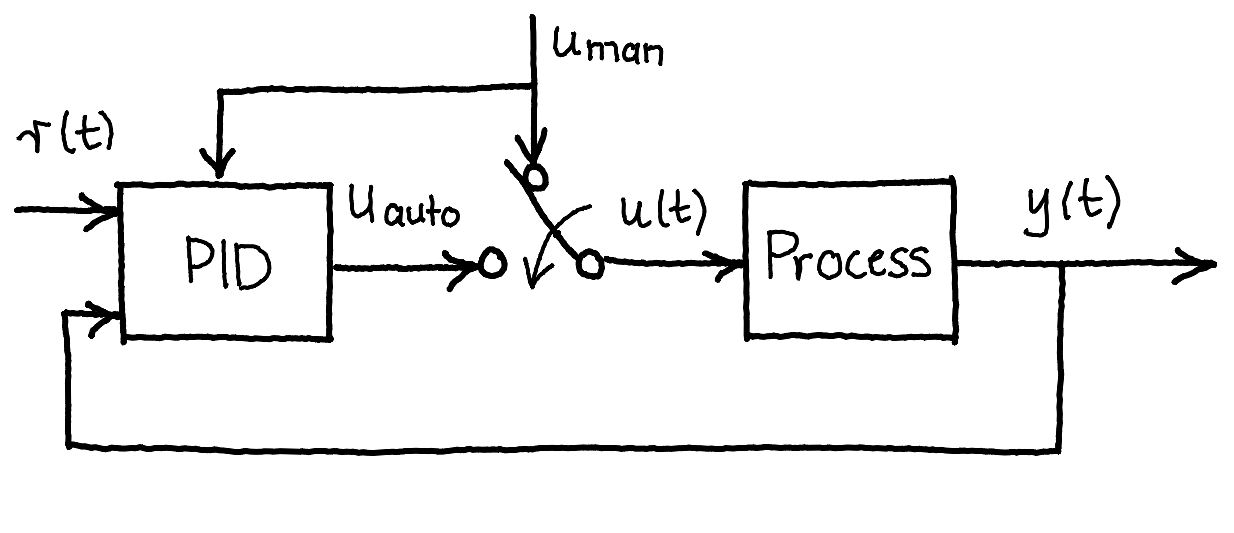}
\caption{Switching between manual and automatic mode.}
\label{fig:ManAuto}
\end{figure}

Since bumpless transfer from manual mode to auto is desired, the manual control signal $u_{\text{man}}$ must enter the controller for this to be implementable; see \cref{fig:ManAuto}.

It is often also desired to have bumpless transfer from automatic mode to manual mode. This can be accomplished outside the controller by setting $u_{\text{man}}$ to $u_{\text{auto}}$ at these mode transitions. Mode switching between man and auto is illustrated in \cref{sec:example1}.\clearpage
\begin{table}
    \renewcommand{\arraystretch}{1.4}
    \centering
       \caption{Bumpless transfer requirements during mode changes and parameter updates for our PID reference implementation. Tracking mode is described in \cref{sec:tracking}.}
    \begin{tabular}{lcc}
    \toprule
         \textbf{Change} & $\mathbf{k_i}$ & \textbf{Bumpless}  \\
        \midrule
         man $\rightarrow$ auto & $k_i \neq 0$ & Yes \\
         auto $\rightarrow$ man & $k_i \neq 0$ & No \\
         man $\rightleftarrows$ auto & $k_i=0$ & No \\
         track $\rightleftarrows$ man & --- & No \\
         track $\rightarrow$ auto & $k_i \neq 0$ & Yes \\
         auto $\rightarrow$ track & $k_i=0$ & No \\
         $k_p$ or $k_d$ update & $k_i \neq 0$ & Yes \\
         $k_p$ or $k_d$ update & $k_i=0$ & No \\
         $k_i$ update to $k_i \neq 0$ & --- & Yes \\
         $k_i$ update to $k_i=0$ & --- & No \\
         \bottomrule
    \end{tabular}
    \label{tab:bumplessReq}
\end{table}

\subsection{Sampling rate and jitter}\label{sec:jitter}
The sampling interval $\Delta t$ is normally set to a constant value so that periodic sampling is obtained. Jitter refers to the deviation from this nominal execution period, whether it was introduced intentionally or accidentally. In most applications, the jitter is small and neglected in the calculations.\clearpage
However, in cases where the jitter is a problem or intentional, for instance in event-based control \rev{\citep{arzen1999}}, $\Delta t$ can be calculated as the actual time elapsed since the last execution. Therefore, the possibility to run a PID controller with jitter compensation is a required functionality for the reference PID controller implementation. Note that a large change in the sampling interval $\Delta t$ also affects the \rev{saturation limits \lstinline{usmin} and \lstinline{usmax} through the rate limits defined by the parameters} $\dot{u}_{\min}$ and $\dot{u}_{\max}$.
\vspace*{-1em}
\subsection{Process variable and setpoint filtering}\label{sec:filter}
The process variable $y(t)$ is often corrupted by high-frequency noise or quantisation effects, which may cause undesired variations in the control signal $u(t)$. These variations are to some extent caused by the proportional term, $u_p(t)$, but more importantly by the derivative term, $u_d(t)$. The integral term $u_i(t)$ averages the noise across samples, and thus serves as a low-pass filter, blocking noise propagation.
Our reference implementation uses the backward difference approximation
\begin{equation}
\frac{dy}{dt} \approx \dfrac{y(t)-y(t-\Delta t)}{\Delta t},
\label{eq:derapprox}
\end{equation}
with $dr/dt$ approximated in the same way, and used in controllers with setpoint derivative weight $c\neq 0$, as described in \cref{sec:setpoint}.
To illustrate noise amplification of the derivative, let $n(t)$ be the additive noise term in $y(t)$. This means that while we consider $y(t)$ to be the process variable, the true, and unknown, process variable is in fact $y(t)-n(t)$. Letting $\Delta n = n(t)-n(t-\Delta t)$ we thus have a noise amplification of $\Delta n / \Delta t$, which is inversely proportional to the sampling interval. For this reason, it is often necessary in practice to low-pass filter the noisy process variable $y(t)$ before computing the finite derivative approximation. This becomes increasingly important as the sampling interval $\Delta t $ decreases.
Similarly, while a true step function exhibits an unbounded derivative, a setpoint step of size $\Delta r$ results in a spike $\Delta r/\Delta t$ in our derivative approximation \cref{eq:derapprox}.
To limit noise amplification, which may otherwise result in unnecessary actuator wear or excite resonant modes in the dynamics of the process, it is customary to low-pass filter the process variable. In situations where a setpoint weight $c>0$ is used in combination with possible abrupt (step) changes in the setpoint $r(t)$, it might also be adequate to low-pass filter $r(t)$.
While a first-order low-pass filter is most often used in existing implementations and textbooks alike, a second-order filter is \rev{often} advisable. This is because the latter results in derivative term noise amplification gain asymptotically approaching zero for high frequencies. This desired property is commonly referred to as high-frequency roll-off \citep{astrom+2006}. This strongly motivates the use of second-order filters, although they introduce slightly more phase-loss than their first-order counterparts.
For our reference implementation, we introduce a simple filter that approximates either the first-order filter with transfer function
\begin{equation}
F_1(s)=\dfrac{1}{sT_f+1},
\label{eq:F1}
\end{equation}
or the second-order filter with transfer function
\begin{equation}
F_2(s)=\dfrac{1}{(sT_f+1)^2},
\label{eq:F2}
\end{equation}
in discrete time, with a sampling interval $\Delta t$. The parameter $T_f$ is the filter time constant, meaning that the magnitude of the filter transfer function has its cutoff at $\omega_c=T_f^{-1}$, and thus effectively attenuates (noise) components in $y(t)$ with angular frequencies larger than $\omega_c$.
We use a discretisation that approximates zero-order-hold (ZOH) well, while being computationally beneficial, and allowing for a filter with bumpless output at changes in either of the filter time constant $T_f$ or the sampling interval $\Delta t$. It is based on a cascaded Tustin approximation, further explained in \cref{sec:filterapp}. The first-order filter \cref{eq:F1} then results in an update equation
\begin{subequations}\label{eq:tustinF}
\begin{equation}
y_f(t)=y_f(t-\Delta t)+a\big(y(t)-y_f(t-\Delta t)\big),
\label{eq:dt_filter}
\end{equation}
with
\begin{equation}
a=\frac{\Delta t}{T_f+\Delta t/2},
\label{eq:a}
\end{equation}
\end{subequations}
where $y$ is the filter input and $y_f$ is the filter output. The second-order filter \cref{eq:F2} is approximated by cascading two first-order filters, as shown in \cref{code:filter}.
The incremental form of \cref{eq:dt_filter} allows for bumpless changes in $T_f$ and $\Delta t$ during operation. Note that there is no formal requirement that the filter run with the same sampling interval as the controller. Noise filtering is illustrated in the example given in \cref{sec:example5}.
\section{The combined form PID controller}\label{sec:CombinedPID}
\begin{figure}[p]
\centering
\subfloat[Schematic representation of \cref{code:control}.]{%
  \includegraphics[width=\textwidth,height=0.45\textheight,keepaspectratio]{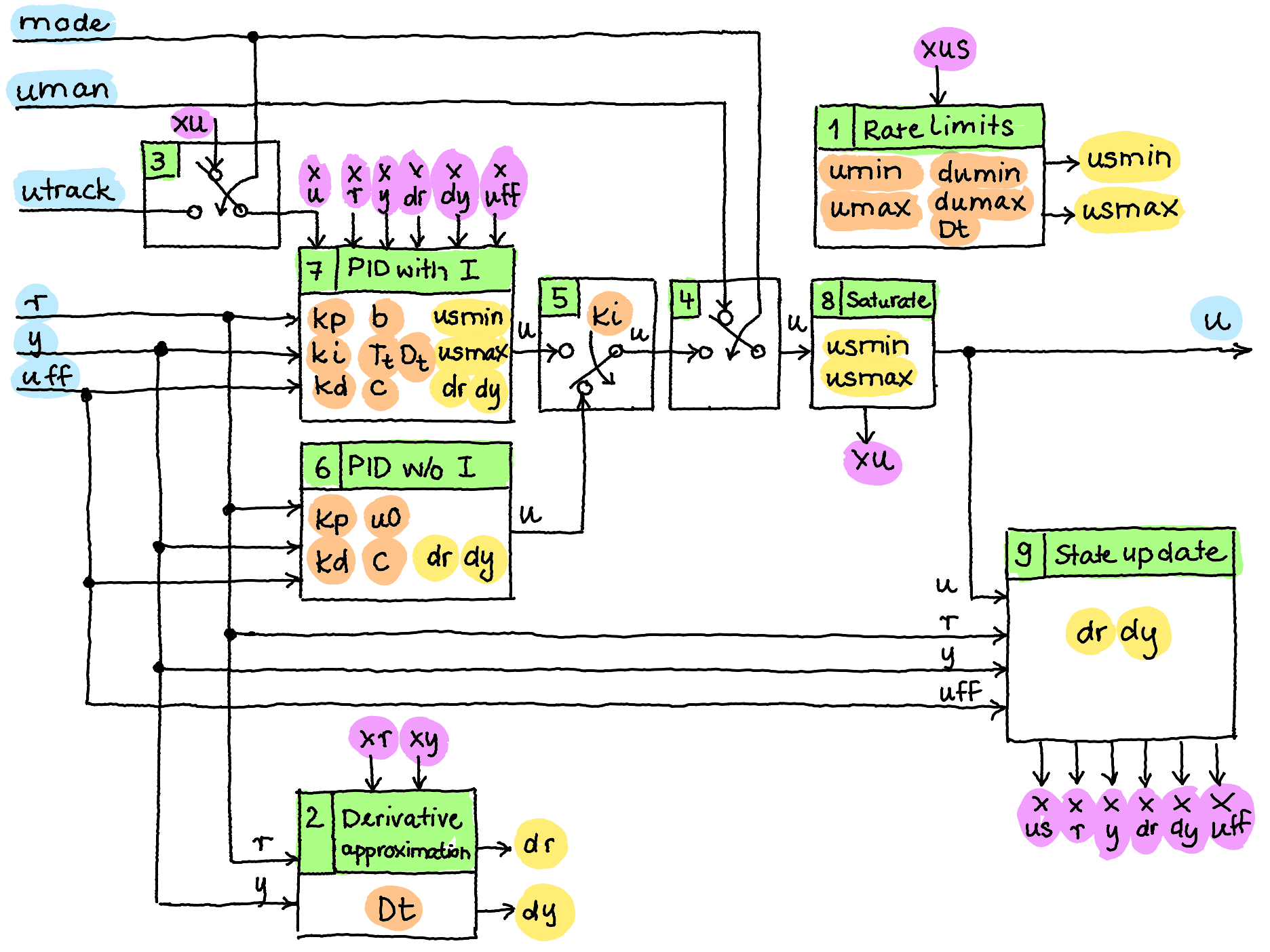}%
  \label{fig:controlschematic1}%
}\par
\vspace{3em}
\subfloat[Schematic representation of the ``PID with I'' block in \cref{fig:controlschematic1}.]{%
  \includegraphics[width=\textwidth,height=0.45\textheight,keepaspectratio]{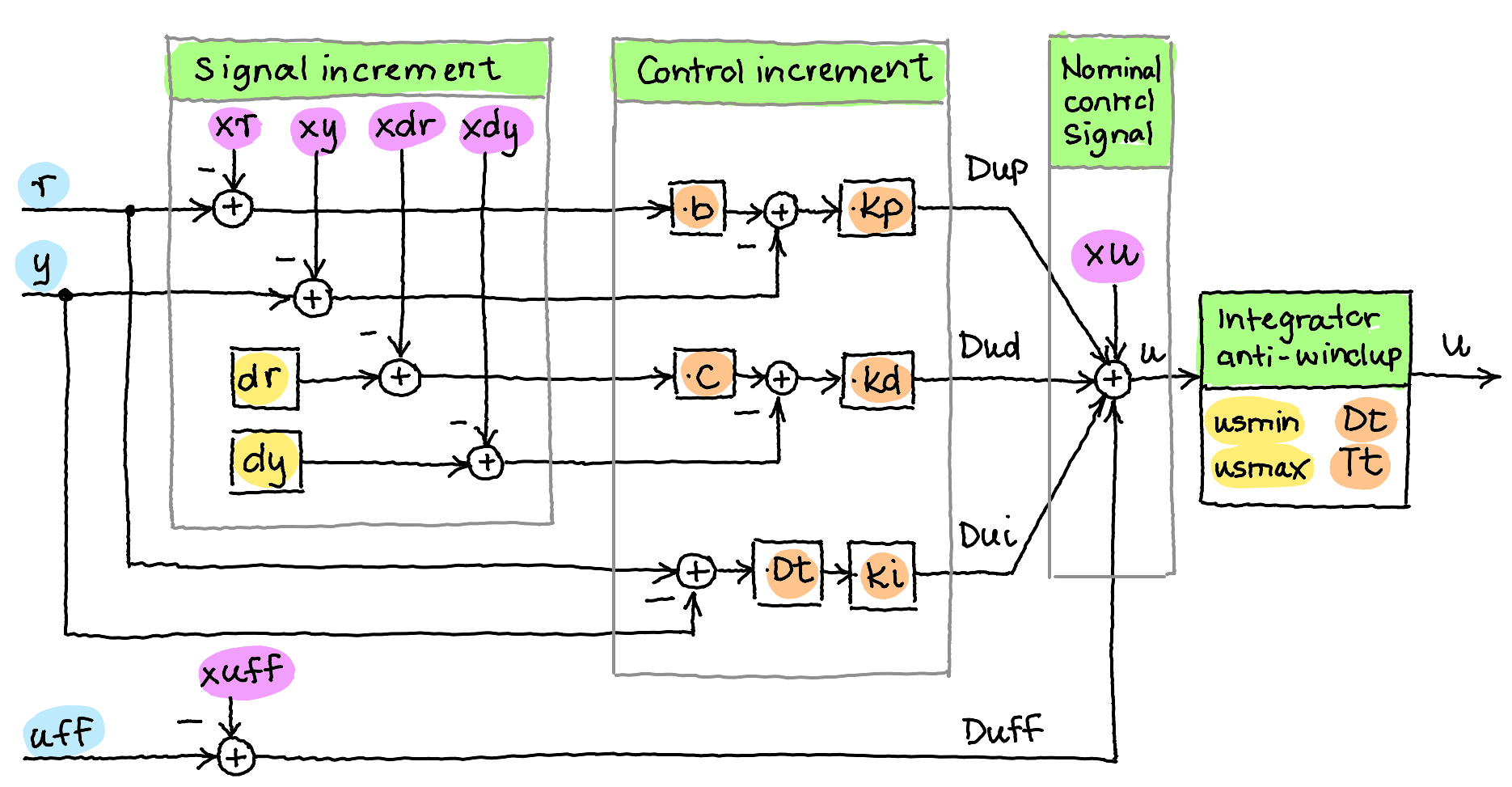}%
  \label{fig:controlschematic2}%
}
\caption{Schematic overview of the combined-form PID controller. Block numbers reflect the order of appearance in our implementation and correspond to the numbered subsections referencing the relevant code lines in \cref{code:control}.}
\label{fig:controlschematic}
\end{figure}

In this section, we propose our reference implementation that implements all functionalities and features described in \cref{sec:features}, comprising the combined form controller of \cref{code:control}, and associated signal filter of \cref{code:filter}.
The code is both generic and general. For example, if a PID controller will never be used in tracking mode, all tracking-specific code can be removed. Similarly, the code can be simplified if, for example, no derivative action or in other words a PI controller, is required.
The term `combined' comes from the controller making use of both the positional and incremental forms, described previously in \crefrange{sec:positional}{sec:incvcpos}. The essential idea is to use the incremental form whenever the controller has integral action. This ensures bumpless behaviour at parameter and mode changes, and facilitates tracking of an exogenous control signal. However, for controllers without integral action, the positional form is needed to ensure the desired response in the control signal. 
In the following sub-sections, we cover features of the proposed implementation to deal with the practical problems described in \cref{sec:features}. The pseudo-code language we use is meant to be generic and focus on algorithmic aspects. For an in-depth explanation of it we refer to \cref{sec:language}.
In the following, we go through our proposed implementation of the combined form PID controller (\cref{code:control}), and the signal filter (\cref{code:filter}) that was proposed in \cref{sec:filter}. We also propose an implementation of the alternative anti-windup strategy that was introduced in \cref{sec:windup} (\cref{code:anti-windup}).
With the exception of the filter and alternative anti-windup strategy, we will use the schematic drawing of \cref{fig:controlschematic} as a basis for introducing the code. The numbered blocks indicate the order in which the corresponding functionality appears in the implementation. Each block is described in a subsection (4.$x$) correspondingly numbered with $x$ referring to the block number in \cref{fig:controlschematic1}. The core part of the implementation, including PID with the integral term and anti-windup techniques, is summarised in \cref{sec:code-integral} and depicted in \cref{fig:controlschematic2}.\clearpage

\lstdefinestyle{controlcolors}{
  literate=
  {b=}{{{\color{black}b=}}}2
  {;}{{\textcolor{lightgray}{;}}}{1}
  {...}{{\textcolor{lightgray}{...}}}{1}
  {[}{{\textcolor{lightgray}{[}}}{1}
  {]}{{\textcolor{lightgray}{]}}}{1}
}
\clearpage
\noindent\makebox[\linewidth][c]{
\begin{minipage}[t]{0.57\linewidth}
\begin{lstlisting}[
  style=controlcolors,
  caption={Combined form PID controller.},
  label={code:control}
]
function u=control(r,y,uff,uman,utrack,mode)|\label{line:controlfunction}\nonum|  
  % 1 Rate limits |\num|
  usmin=max(umin,xus+Dt*dumin) |\label{line:usmin}|
  usmax=min(umax,xus+Dt*dumax)  |\label{line:usmax}\nonum|

  % 2 Derivative approximation |\numdecr|
  dr=(r-xr)/Dt |\label{line:dr}|
  dy=(y-xy)/Dt |\label{line:dy}\nonum|

  %   3 Tracking mode |\numdecr|
  if mode==TRACK |\label{line:iftrack}|
    xu=utrack |\label{line:tracku}|
  end |\label{line:endtrack}\nonum|
   % 4 Manual mode |\num|
  if mode==MAN |\label{line:ifman}|
    u=uman |\label{line:uman}\nonum|
    % 5 Integral switch |\num|
  elseif ki==0 |\label{line:ifint}\nonum|
    % 6 PID without I |\num|
    u=u0+kp*(r-y)+kd*(c*dr-dy)+uff|\label{line:upos}|
  else  |\label{line:intstart}\nonum| 
    % 7 PID with I 
    % Signal increment |\num|
    Dr=r-xr |\label{line:Dr}|
    Dy=y-xy |\label{line:Dy}|
    Ddr=dr-xdr |\label{line:Ddr}|
    Ddy=dy-xdy  |\label{line:Ddy}\nonum|

    % Control increment  |\numdecr|
    Dup=kp*(b*Dr-Dy) |\label{line:Dup}|
    Dui=ki*Dt*(r-y) |\label{line:Dui}|
    Dud=kd*(c*Ddr-Ddy) |\label{line:Dud}|     
    Duff=uff-xuff  |\label{line:Duff}\nonum|

    % Nominal control signal  |\numdecr|
    u=xu+Dup+Dui+Dud+Duff  |\label{line:unom}\nonum|

    % Integrator anti-windup  |\numdecr|
    us=max(min(u,usmax),usmin) |\label{line:usminmax}|
    u=u-Dt/Tt*(u-us) |\label{line:uaw}|
  end  |\label{line:endmain}\nonum|

  % 8 Saturate  |\numdecr|
  xu=u |\label{line:xu}|
  u=max(min(u,usmax),usmin) |\label{line:uminmax}\nonum|
  
  % 9 State update  |\numdecr|
  xus=u  |\label{line:xus}|
  xr=r |\label{line:xr}|
  xdr=dr |\label{line:xdr}|
  xy=y |\label{line:xy}|
  xdy=dy |\label{line:xdy}|
  xuff=uff |\label{line:xuff}|
end |\label{line:controlend}|
\end{lstlisting}
\end{minipage}\hspace{0.04\linewidth}
\begin{minipage}[t]{0.52\linewidth}
\begin{lstlisting}[
  caption={Second-order signal filter.},
  label={code:filter}
]
function xf2=filter(y) |\label{line:filterfunction}| |\nonum|
  % Filter parameter |\num|
  a=Dt/(Tf+0.5*Dt) |\label{line:a}\nonum|

  % State update |\numdecr|
  xf1=xf1+a*(y-xf1) |\label{line:x1}|
  xf2=xf2+a*(xf1-xf2) |\label{line:x2}|
end |\label{line:filterend}|
\end{lstlisting}
\begin{lstlisting}[
  caption={Combined integration anti-windup. Can replace \Line{uaw} in \cref{code:control}.},
  label={code:anti-windup}
]
if Dui*(u-us)>0 |\label{line:intcond}|
  u=u-sign(u-us)*min(abs(Dui),abs(u-us))|\label{line:condint2}|
end |\label{line:awend}|
u=u-min(Dt/Tt,1)*(u-us) |\label{line:uaw2}|
\end{lstlisting}
\end{minipage}
}
\clearpage

In \cref{code:control}, the \lstinline{control} function itself is invoked on \line{controlfunction}, and takes as its arguments:
\begin{itemize}
\item setpoint \lstinline{r},
\item process variable \lstinline{y},
\item feed-forward control signal \lstinline{uff},
\item manual control signal \lstinline{uman},
\item tracking control signal \lstinline{utrack},
\item operating mode \lstinline{mode}.
\end{itemize}
The process variable \lstinline{y} and setpoint \lstinline{r} are assumed to be adequately filtered, as discussed in \cref{sec:filter}, using for example our proposed filter implementation of \cref{sec:code-filter}. The mode variable \lstinline{mode} has three defined values: automatic mode \lstinline{AUTO}, manual mode \lstinline{MAN}, and external signal tracking mode \lstinline{TRACK}. The controller is in automatic mode whenever \lstinline{mode} is neither \lstinline{MAN} nor \lstinline{TRACK}. Encoding of the mode is further discussed in \cref{sec:ControlMode}. If no feed-forward action is desired, \lstinline{uff} is simply set to zero. On a similar note, the values of \lstinline{uman} and \lstinline{utrack} are relevant when the controller is in manual and tracking mode, respectively.
The implementation also consists of both states and parameters. The parameters, e.g. controller gains and limits, are set externally by some language-dependent unspecified mechanism, further discussed in \cref{sec:closedloopimp}. States are used to store controller values between executions. Complete lists of all arguments, parameters, and states are found in \cref{sec:nomenclature}.

\subsection{Rate limits
}\label{sec:code-ratelim}
The parameters \lstinline{umin} and \lstinline{umax} define the interval into which the control signal is saturated, cf. \cref{sec:code-saturation}. The parameters \lstinline{dumin} and \lstinline{dumax} define a lower and upper rate limit for the change of the control signal. They are used together with the previous value of the saturated control signal, \lstinline{xus} and the past sampling interval \lstinline{Dt} to compute corresponding saturation bounds \lstinline{xus+Dt*dumin} and \lstinline{xus+Dt*dumax} (see \cref{sec:ulims}).
Note that it is important that the previous saturated control signal \lstinline{xus}, and not the previous nominal control signal \lstinline{xu} is used to produce a correct rate limitation.
The actual saturation limits \lstinline{usmin} and \lstinline{usmax} \rev{are computed as the more conservative of the absolute limits defined by the parameters \lstinline{umin} and \lstinline{umax}, and those imposed by the rate limit parameters \lstinline{dumin} and \lstinline{dumax}}, as coded on \lines{usmin}{usmax}.
\subsection{Derivative approximation
}\label{sec:code-derivatives}
The derivatives used by controllers with derivative action are approximated using finite differences, as explained in \cref{sec:sampling}. The difference between the current and previous process variable value \lstinline{y-xy} is divided by the execution interval duration \lstinline{Dt}, to produce the finite difference approximation \lstinline{dy} on \line{dy}. The setpoint derivative approximation \lstinline{dr} is computed in the same way on \line{dr}.
To avoid undesired jumps and noise amplification in the control signal, it is important that the process variable is adequately filtered, as described in \cref{sec:filter}. In the case of nonzero setpoint weight \lstinline{c}, the setpoint must also be adequately filtered.
\subsection{Tracking mode
}\label{sec:code-track}
Tracking of external signals, as explained in \cref{sec:tracking} is implemented to overwrite the previous output \lstinline{xu} with the signal to be tracked, \lstinline{utrack} on \line{tracku}. Then the control signal increment is computed as normal, cf. \cref{sec:code-integral}. Tracking mode is active when \lstinline{mode==TRACK}.
Note that tracking is only meaningful when there is integral action. In the case of no integral action, tracking mode and manual mode behave identically.
In our implementation, feed-forward and integrator anti-windup behave the same way in tracking mode as in automatic mode.
\subsection{Manual mode
}\label{sec:code-man}
The controller is in manual mode when \lstinline{mode==MAN}. In manual mode, the nominal control signal \lstinline{u} takes on the externally provided value \lstinline{uman}, as expressed on \line{uman}. The manual mode control signal is limited as explained in \cref{sec:code-ratelim}. Feed-forward action is disabled in manual mode. Due to the incremental form when we have integral action, the bumpless mode switches required in \cref{sec:manauto} are automatically obtained.
\vspace*{-1em}
\subsection{Integral switch
}\label{sec:code-formselect}
The combined form switches between positional and incremental form based on whether the controller utilises integral action, as specified in \cref{sec:PandPD}. This is achieved by checking for \lstinline{ki==0} on \line{ifint}. If the condition is true, positional form is employed, as explained in \cref{sec:code-nointegral}. Otherwise, the \lstinline{else} block starting on \line{intstart} is executed, implementing incremental form as explained in \cref{sec:code-integral}.
\subsection{PID without integral action
}\label{sec:code-nointegral}
In the absence of integral action (\lstinline{ki==0}), the positional form control signal is computed according to \line{upos}. The constant bias term \lstinline{u0} can be used to zero-offset the control signal to match the stationary operating point of consideration. \rev{Setpoint weighting with weight parameter \lstinline{c} is used for the derivative term, but for the proportional term the weight is fixed to \lstinline{b} =1,} as explained in \cref{sec:setpoint}.
Feed-forward is enabled, and the control signal is saturated according to \cref{sec:code-ratelim}, but since there is no integral action, no integrator anti-windup is applied.
Parameter and mode changes are not bumpless, as there is no integrator state that can be shifted to compensate for bumps.
\subsection{PID with integral action
}\label{sec:code-integral}
With integral action (\lstinline{ki} not zero), an incremental form PID control law is applied, which is summarised in \Cref{fig:controlschematic2} according to the code \lines{Dr}{uaw}.
The setpoint and process variable increments \lstinline{Dr} and \lstinline{Dy} are computed on \lines{Dr}{Dy}, followed by increments of their derivative approximations (see \cref{sec:code-derivatives}) on \lines{Ddr}{Ddy}. Taking the process variable as example, the increment is simply computed as the current value \lstinline{y} minus the previous value \lstinline{xy} stored in the previous execution of the \lstinline{control} function.
The computed signal increments are then used to compute increments of the individual control signal terms on \lines{Dup}{Duff}. Since the derivatives are calculated from the actual time between executions, \lstinline{Dt}, and the integral increment also uses \lstinline{Dt}, we compensate for possible jitter according to \cref{sec:jitter}.\clearpage
Weight parameters \lstinline{b} and \lstinline{c} are used for the proportional and derivative terms, respectively, to account for setpoint weighting as described in \cref{sec:setpoint}. The rectangular integral approximation and finite difference derivative approximation utilised in computing \lstinline{Dui} and \lstinline{Dud} are described in \cref{sec:sampling}.
The nominal (i.e., not yet saturated) control signal \lstinline{u} is computed on \line{unom} by adding the increments from \lines{Dup}{Duff} to the previous nominal control signal \lstinline{xu}. Note that the feed-forward signal must enter the signal increment here, as motivated in \cref{sec:FF}.
Note that it is important that it is the nominal previous control signal \lstinline{xu} that is used, and not its saturated counterpart. Otherwise, it is not possible to implement integrator anti-windup strategies other than control signal clamping, as both conditional integration and back calculation rely on the possibility for the control signal to go outside of its unsaturated range.
As explained in \cref{sec:windup}, integrator anti-windup relies on knowing the difference between the nominal control signal \lstinline{u}, and its saturated counterpart that we call \lstinline{us}. To this end, the saturation limits computed on \lines{usmin}{usmax}, as explained in \cref{sec:code-ratelim}, are applied on \line{usminmax} to obtain \lstinline{us}.
The reference implementation of \cref{code:control} implements the back calculation anti-windup strategy on \line{uaw} (with control signal clamping as a special case when \lstinline{Tt} equals \lstinline{Dt}).
A more general anti-windup implementation, that can serve as an in-place replacement for \line{uaw} of \cref{code:control} is provided in \cref{code:anti-windup}. It fulfils the advanced anti-windup behaviours specified at the end of \cref{sec:windup}.
For integrator clamping, it can be used as provided with \lstinline{Tt} set to \lstinline{Inf}, effectively translating \line{uaw2} into \lstinline{u=u}. Alternatively \line{uaw2} can be removed entirely when integrator clamping is desired.
\Lines{intcond}{awend} in \cref{code:anti-windup} have no purpose in control signal clamping, and if no other anti-windup mode needs to be supported, \cref{code:anti-windup} can be replaced in its entirety by \lstinline{u=us}. An even simpler alternative for this case is to move saving of the control signal state (\lstinline{xu=u}) on \line{xu} of \cref{code:control} to after saturating the control signal on \line{usminmax}, and removing the entire integrator anti-windup block on \lines{usminmax}{uaw}; replacing \lstinline{xus} by \lstinline{xu} on \lines{usmin}{usmax}, and removing \line{xus}.
\subsection{Saturate 
}\label{sec:code-saturation}
Upon saving the nominal, unsaturated, control signal as \lstinline{xu} on \line{xu}, the control signal is saturated on \line{uminmax}, as described in \cref{sec:code-ratelim}. It is this saturated control signal that is eventually returned by the \lstinline{control} function.
\subsection{State update
}\label{sec:code-stateupdate}
As a final stage of the \lstinline{control} function, the state variables are updated on \lines{xus}{xuff} (in addition to \lstinline{xu} that was updated on \line{xu} before saturating the control signal).
\subsection{Process variable and setpoint filtering}\label{sec:code-filter}
As discussed in \cref{sec:filter}, the process variable, and sometimes the setpoint, need to be adequately low-pass filtered before being passed to the PID control algorithm of \cref{code:control}. To this end, \cref{code:filter} implements the low-pass filter introduced in \cref{sec:filter} and further explained in \cref{sec:filterapp}.
\Line{a} computes the filter parameter \lstinline{a} from the filter time constant parameter \lstinline{Tf} and execution interval \lstinline{Dt}, as explained in \cref{sec:filterapp}. If the filter runs at a constant sampling interval, there is no need to re-compute \lstinline{a} at each invocation of the \lstinline{filter} function. Instead, \lstinline{a} can then be considered a parameter, computed outside the \lstinline{filter} function and provided in the same fashion as other parameters.\clearpage
The cascaded filter is implemented on \lines{x1}{x2}. If a first-order filter is desired, \line{x2} can be removed, and \lstinline{xf2} changed to \lstinline{xf1} on \line{filterfunction}.
Note that the filter function must be instantiated separately for every signal being filtered, since each instance requires its own state variables \lstinline{xf1} and \lstinline{xf2}.
\section{Closed loop implementation} \label{sec:closedloopimp}
The PID control algorithm may be executed in a real-world, run-time environment or in a simulated environment, but always in closed loop. Both scenarios require more code than the function presented in \cref{sec:CombinedPID}. This section first provides PID function execution details including initialisation, default parameters and discussions relating to the control mode. It then provides the code necessary for closed loop implementation as will be used in \cref{sec:basicexamples}.
\subsection{PID function execution}
The PID function call cannot be stand-alone but is embedded in an environment. The environment can take various forms and is usually application specific, but a number of common questions arise relating to the execution: how are states and signals initialised? How are parameters passed to the function?
Answers to these questions are given in the following sections.
\subsubsection{Initialisation} \label{sec:initialisation}
The state variables of the controller, as well as any signal filters, must be initialised to appropriate values before the \lstinline{control} or \lstinline{filter} functions of \cref{code:control} and \cref{code:filter} are invoked for the first time. The controller states are \lstinline{xr}, \lstinline{xu}, \lstinline{xus}, \lstinline{xy}, \lstinline{xdy}, \lstinline{xdr} and \lstinline{xuff}. The latest available signal values can be used to initialise the controller states, so that, for example, \lstinline{xr=r}, etc.
If the controller is started in manual mode, the initialisation of \lstinline{xu} has no practical effect. If, instead, the controller is started in automatic or tracking mode, it is reasonable to initialise \lstinline{xu} to the value that the actuator can be expected to have at startup. For this reason, setting \lstinline{xu=0} and \lstinline{xus=0} is appropriate in many situations. However, other initialisations may be preferable depending on the application context. As with controller parameter tuning, such choices depend more on the specific use case than on the controller implementation itself, and are therefore not discussed further here.
A reasonable initialisation of the low-pass filter is to assign \lstinline{xf1} and \lstinline{xf2} the latest available value of the corresponding unfiltered measurement.
To avoid transient effects, it is advisable to let filters converge to their steady-state before placing the corresponding controller in automatic or tracking mode.
In the closed-loop simulation examples of this paper, the initial variables and subsequently the controller states are all set to zero by default.
\subsubsection{PID parameters}
The parameters, which are {\color{amber}orange} in colour in the code, are listed in \cref{sec:nomenclature}. If parameters are passed by reference, it is important to prevent them from changing while critical parts of the functions are executing. For example, switching to a new pair \lstinline{kp}, \lstinline{ki} should be done in a way that avoids a single invocation of \cref{code:control} using the old \lstinline{kp} value together with the new \lstinline{ki} value. This can be avoided in several ways.\clearpage
One approach is to pass all parameters by value as arguments to the function that uses them. Another approach is to call an update function that reads parameters from memory at the beginning of the function. A third option is to use thread-safety mechanisms such as locks or mutexes to prevent parameter updates while critical sections of the function are executing.
The reference implementation assumes a runtime capable of performing floating-point arithmetic, either natively through a floating-point unit or via emulation. Established methods exist for porting floating-point implementations to fixed-point architectures. We therefore do not delve further into this topic, but refer the interested reader to the introduction provided in \cite{crisp1991}.
One important aspect concerns how signals and parameters are passed to the controller, and how the computed control signal is actuated. In our reference implementation, signals are passed as arguments to the \lstinline{control} and \lstinline{filter} functions (see \cref{code:control} and \cref{code:filter}), while parameters (orange) and state variables (pink) are assumed to be provided through some unspecified mechanism.
It should be noted that the distinction between signals and parameters is not always clear-cut. For example, \lstinline{mode} could just as well be regarded as a parameter. We deliberately refrain from prescribing how values are passed, since this depends on programming language and runtime environment.
In an object-oriented setting, it is natural to define a PID object with \lstinline{control} and \lstinline{filter} methods, together with appropriate constructors for initialisation, as the MATLAB and Python implementations in our GitHub repository \citep{sundstrom+2026}. However, the proposed algorithms can equally well be implemented without object-oriented constructs. For this reason, we focus on the algorithmic structure rather than on software architectural details.

\subsubsection{Control mode} \label{sec:ControlMode}
The key function of the control mode is to allow the PID user or operator to switch the controller on or off, which is labelled automatic (\lstinline{AUTO}) and manual (\lstinline{MAN}), respectively. As described in \cref{sec:manauto}, the controller can be in tracking mode when the calculated control signal is not applied but the controller tracks another signal.
There is a risk that the controller defaults to automatic mode on \line{intstart} if \lstinline{mode} holds a value other than \lstinline{MAN} or \lstinline{TRACK}. This risk can be eliminated by introducing the fourth mode `Disabled'. The `Disabled' mode means that the control signal $u$ is neither calculated nor applied.
\subsection{Basic closed-loop simulation}
To illustrate simulation use, \cref{code:basicloop} shows the code for a basic control loop. The simulation codes for this and for the examples in \cref{sec:basicexamples} are available on GitHub.

The PID controller parameters, saturation limits, rate limits, sampling interval, and filter time constant are set on \lines{blparams}{blparams2}, and the PID controller is created from those parameters on \line{blpid}. The initial values for the different signals are given on \line{blsignals}, also setting the controller mode to \lstinline{AUTO}. These values are used on \line{blinit} to initialise the PID controller states by calling the \lstinline{initialisation} function according to the ideas described in \cref{sec:initialisation}.
 The control loop for a given simulation time \lstinline{Tsim} is defined on \lines{blloop}{blloopend}. Inside the loop, \line{blsensor} simulates the process dynamics, a setpoint change is configured on \lines{blsp}{blspend}, the process output is filtered on \line{blfilter} using the \lstinline{filter} function, and the controller action is calculated on \line{blcontrol} by calling the \lstinline{control} function. The calculated control signal is sent to the plant on \line{blset}, and the loop waits for the next sampling instant on \line{blwait}. Unless otherwise stated, the initial conditions defined on \lines{blparams}{blinit} of \cref{code:basicloop} are used as default values for all examples presented in \cref{sec:basicexamples}, and only the specific modifications are described in the corresponding sections.
\clearpage
\begin{lstlisting}[
  style=controlcolors,
  caption={Basic control loop.},
  label={code:basicloop}
]
kp=1.0 ki=0.5 kd=0.1 b=1 c=0 umin=0 umax=100 |\label{line:blparams}|
dumin=-10 dumax=10 u0=0 Tt=0.1 Tf=0.1 Dt=0.1 |\label{line:blparams2}\nonum|
% Creating PID instance with parameter values |\num|
pid1=PID(kp,ki,kd,b,c,u0,umin,umax,dumin,dumax,Tt,Tf,Dt) |\label{line:blpid}\nonum| 

%  PID states and signals |\numdecr|
r=0 y=0 dr=0 dy=0 u=0 uff=0 us=0 uman=0 utrack=0 mode="AUTO" |\label{line:blsignals}\nonum|
% Initialise PID states |\num|
pid1.initialisation(r,y,dr,dy,uff,u,us) |\label{line:blinit}\nonum|

% Execution Loop |\numdecr|
while t < Tsim |\label{line:blloop}\nonum| 

    % Set-point change |\numdecr|
    if t>=2 |\label{line:blsp}|
        r=1
    end |\label{line:blspend}\nonum| 

    % Read from sensor   |\numdecr|
    y=read_process_variable() |\label{line:blsensor}\nonum| 
        
    % Filtering process output |\numdecr|
    yf=pid1.filter(y,Tf,Dt) |\label{line:blfilter}\nonum|

    % PID control signal calculation |\numdecr|
    u=pid1.control(r,yf,uff,uman,utrack,mode) |\label{line:blcontrol}\nonum|

    % Set control signal   |\numdecr|
    set_control(u) |\label{line:blset}\nonum|

    % Wait for next iteration |\numdecr|
    wait(Dt) |\label{line:blwait}|
end|\label{line:blloopend}|
\end{lstlisting}
\section{Simulation examples}\label{sec:basicexamples}
This section presents seven simulation examples to show how the proposed code handles the different implementation features discussed in \Cref{sec:features}. The first three examples deal with switching between automatic and manual mode, switching the integral action on and off, and showing the effect of rate limitation, respectively. The fourth example is focused on analysing the combination of the PID controller with a feed-forward compensator and different solutions to treat the control signal saturation problem. The measurement noise filtering capabilities are presented in the fifth example. Finally, the last two examples demonstrate the use of the basic PID functionality for the cases that involve tracking a control signal, in particular gain scheduling and selector control.
For most of the examples, a first-order plus dead-time process model with transfer function
\begin{equation}
    P(s) = \frac{k}{1+sT}e^{-sL}=\frac{1}{1+s}e^{-0.5s}
    \label{eq:P}
\end{equation}
is used and controlled by a PI-controller tuned using the Lambda method \rev{\citep{dahlin1968}}, with
$\lambda=T=1$ and sampling interval $0.1$.\clearpage
\noindent This results in the controller
\begin{equation}
    C(s) = K\left(1+\frac{1}{sT_i}\right)=0.667\left(1+\frac{1}{s}\right) = k_p+k_i\frac{1}{s}=0.667 + 0.667\frac{1}{s}.
    \label{eq:C}
\end{equation}
When different process dynamics or controller parameters are required for some particular example, new information will be provided.
\subsection{Example 1: Man/Auto switching}\label{sec:example1}
\begin{figure}[ht]
    \centering \includegraphics[width=0.7\linewidth]{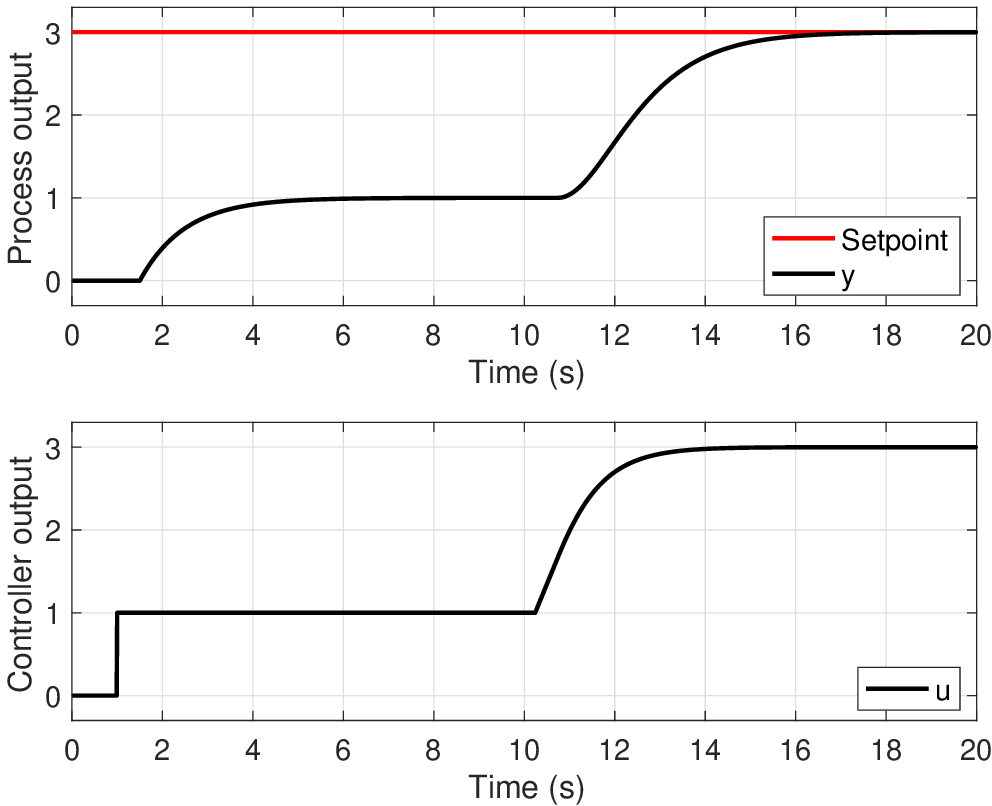}
    \caption{Switching from manual to automatic mode in Example~1.}
    \label{fig:example1}
\end{figure}

In the proposed simulation, the system starts in manual mode at the beginning of the simulation and then switches to automatic mode in the middle of the simulation time.
\Cref{fig:example1} shows a simulation for this example in which the system starts in manual mode from $t=0$ to $t=10$, with a step control signal change from $u_{\text{man}}=0$ to $u_{\text{man}}=1$ at $t=1$, and switches from manual to automatic mode at time $t=10$. The setpoint is fixed to $r=3$ from the beginning of the simulation.
As observed, the switch between manual and automatic mode is performed properly, and there is no bump at the switching time.
\Cref{code:example1} shows the control loop code for this example. Before simulating the control loop, the following changes must be made in the initialisation section for parameters and states: \lstinline{kp=0.667}, \lstinline{ki=0.667}, and \lstinline{kd=0.0}, \lstinline{mode=MAN}, and \lstinline{r=3}. As observed, the step control signal change for manual mode is done on \lines{ex1man}{ex1manend} setting \lstinline{uman=1}, while the switch between manual and automatic modes is performed on \lines{ex1auto}{ex1autoend} setting \lstinline{mode=AUTO}.
\clearpage
\begin{lstlisting}[
  caption={Example 1 code.},
  label={code:example1}
]
while t < Tsim |\nonum|
    % Simulation of process dynamics |\num| 
    y=process_simulation(u,Dt) |\nonum|
        
    % Step change on manual input signal |\numdecr| 
    if t>=1 |\label{line:ex1man}|
        uman=1;
    end |\label{line:ex1manend}\nonum|     

    % Switch to AUTO |\numdecr| 
    if t>=10 |\label{line:ex1auto}|
        mode="AUTO"
    end |\label{line:ex1autoend}\nonum|

    % PID control signal calculation |\numdecr| 
    u=pid1.control(r,y,uff,uman,utrack,mode); |\nonum|

    % Simulation time update |\numdecr| 
    t=t+Dt
end
\end{lstlisting}
\subsection{Example 2: P- or PD-control and rate limitation}\label{sec:example2}
\begin{figure}[!ht]
    \centering
    \includegraphics[width=0.7\linewidth]{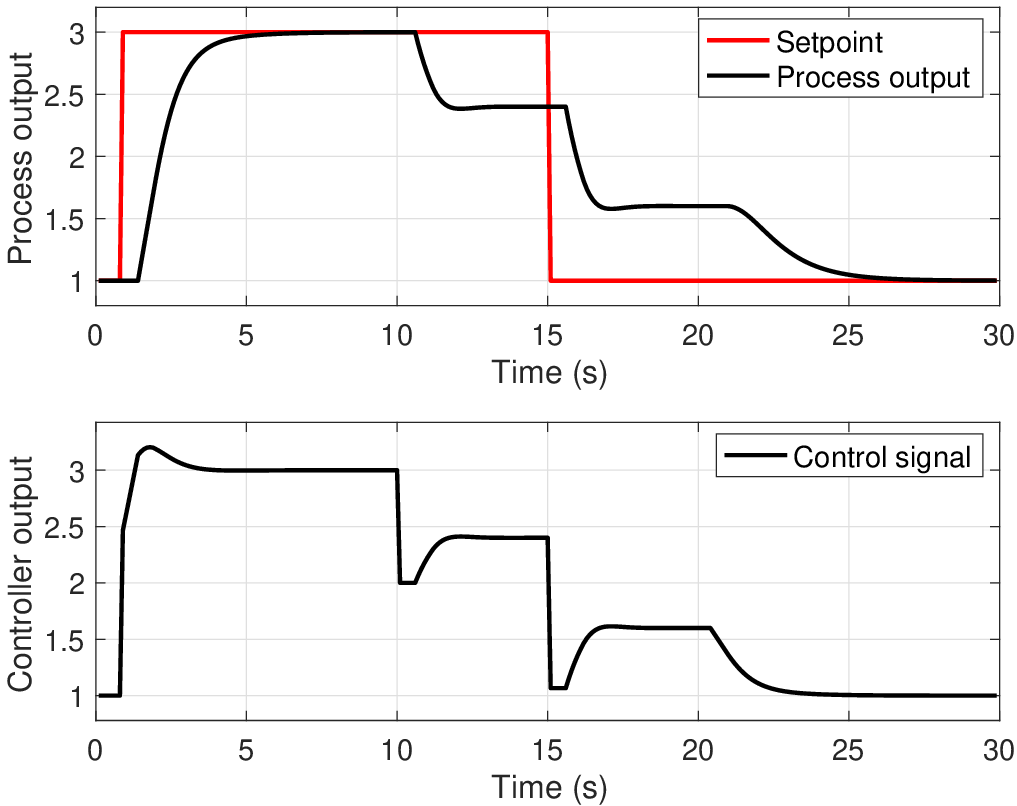}
    \caption{Switching between PI- and P-control in Example~2.}
    \label{fig:example2}
\end{figure}

\begin{figure}[!ht]
    \centering
    \includegraphics[width=0.7\linewidth]{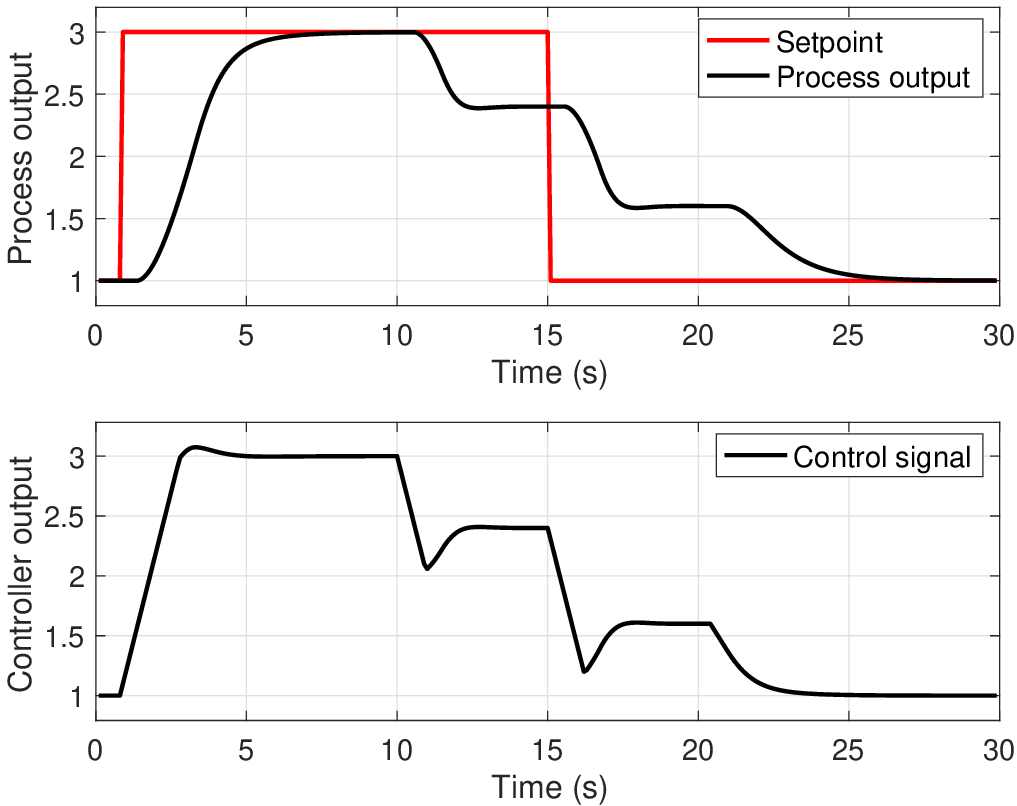}
    \caption{\rev{Rate limitation effect in Example~2 for \lstinline{dumin=-1} and \lstinline{dumax=1}.}}
    \label{fig:example2_rate_limitation}
\end{figure}

This example shows how the proposed code can be used to implement P- or PD-controllers, that is, controllers without integral action. The rate limitation function is also illustrated. We use the process and controller given by \cref{eq:P} and \cref{eq:C}.
For the P-controller the integral gain is set to $k_i=0$ and the bias term $u_0$ is introduced with $u_0=2$. \Cref{fig:example2} shows the simulation results for this example.
\clearpage
The simulation starts using the PI-controller for an operating point given by $y=1$ and $u=1$, and with a setpoint change from $r=1$ to $r=3$ at $t=1$. At time $t=10$, the controller is switched to a P-controller by setting $k_i=0$. Since the control error is zero at time $t=10$, the control signal will jump down to the bias term $u_0=2$, and stay there until the process delay has elapsed. The P-controller will then control the process variable to a new position, but since there is no integral action there will be a control error.
At time $t=15$, another setpoint change is made, from $r=3$ to $r=1$. The P-controller behaves as expected, and the two setpoint responses are similar, showing that the transitions from PI to P work well.
Finally, at time $t=21$ we switch back to a PI-controller and the control error is eliminated, obtaining a bumpless transfer response.
For this example, the simulation code is not shown since it is similar to the code used in \cref{code:example1}. The manual part in \cref{code:example1} is removed, the variable \lstinline{ki} is set to \lstinline{ki=0} at time $t=10$ to introduce the P-controller, and it is set to \lstinline{ki=0.667} at $t=21$ to go back to the PI-controller again. At $t=1$ and $t=15$ setpoint changes are introduced by setting \lstinline{r=3} and \lstinline{r=1}, respectively, in a similar way as shown in \cref{code:basicloop} on \lines{blsp}{blspend}. Moreover, the following changes are required at the initialisation section: \lstinline{y=1.0}, \lstinline{u=1.0}, \lstinline{u0=2.0}, \lstinline{r=1.0}, and \lstinline{mode=AUTO}.
This example is also used to evaluate the rate limitation option. \Cref{fig:example2_rate_limitation} shows the same simulation results presented in \cref{fig:example2} but for a rate limit of 1. As observed at time instants $t=1$, $t=10$, and $t=15$, the changes of the control signal are now limited and thus a slower response is obtained according to the rate limitation imposed. For this simulation, rate limitation variables must be modified in the code by setting \lstinline{dumin=-1} and \lstinline{dumax=1} in the initialisation section.
\vspace*{-2em}
\subsection{Example 3: Setpoint handling}\label{sec:example3}
\begin{figure}[ht]
    \centering
    \includegraphics[width=0.7\linewidth]{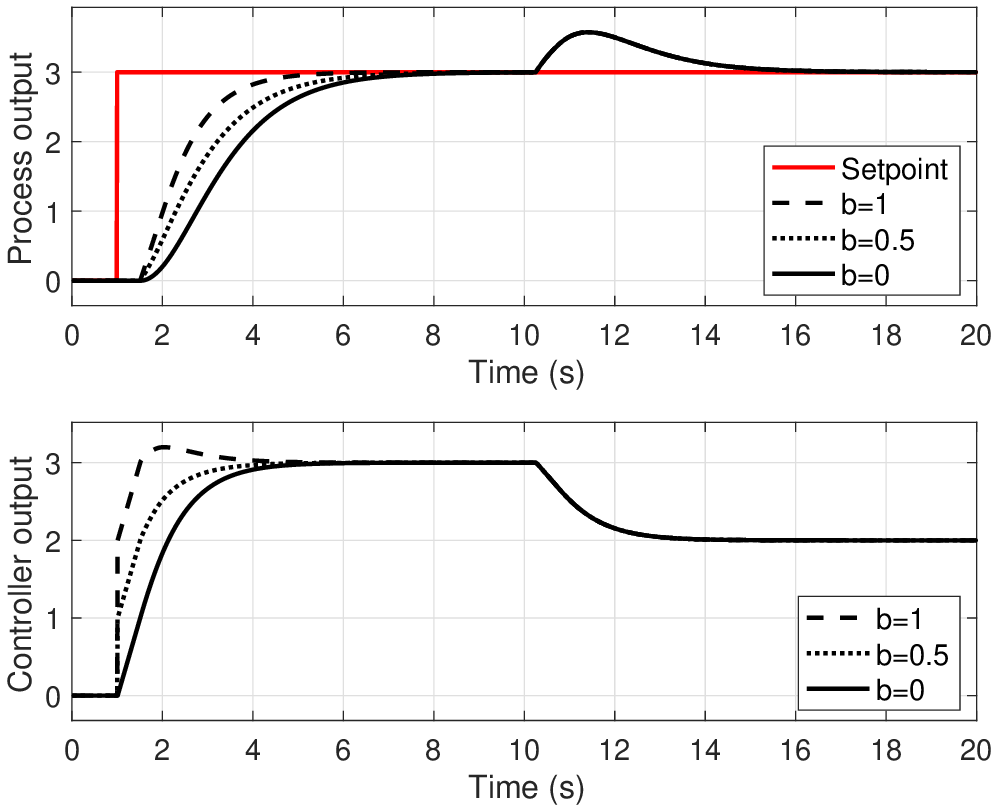}
    \caption{Setpoint handling in Example~3.}
    \label{fig:example3}
\end{figure}

As discussed in \cref{sec:setpoint}, the code proposed in this paper includes the capabilities for setpoint handling to obtain a smoother control signal when strong changes of setpoints (like step changes) are required.
To show this idea, in this example we use the process and controller given by \cref{eq:P} and \cref{eq:C}.\clearpage
As the aim of this example is to show the effect of the $b$ parameter on the responses of the control system, \cref{fig:example3} shows the simulation results for the cases with $b=0$, $b=0.5$, and $b=1$. As observed, the smoother control signal is obtained for the case with $b=0$ and the more aggressive control signal for $b=1$. The $b$ factor will of course also influence the process variable response, where the slower and faster responses are obtained for $b=0$ and $b=1$, respectively.
\Cref{fig:example3} also shows the effect on the disturbance rejection for a step-like disturbance signal of amplitude 1 added to the control signal at time $t=10$. As expected, identical responses are obtained for any value of $b$, as the setpoint handling is decoupled from the disturbance rejection problem.
For this example, the simulation code is not shown as it is the same code as used in \cref{code:example1} skipping the manual part at the beginning of the simulation and using different values for the $b$ parameter.
\subsection{Example 4: Feed-forward and saturation}\label{sec:example4}
\begin{figure}[H]
    \centering
    \includegraphics[width=0.7\linewidth]{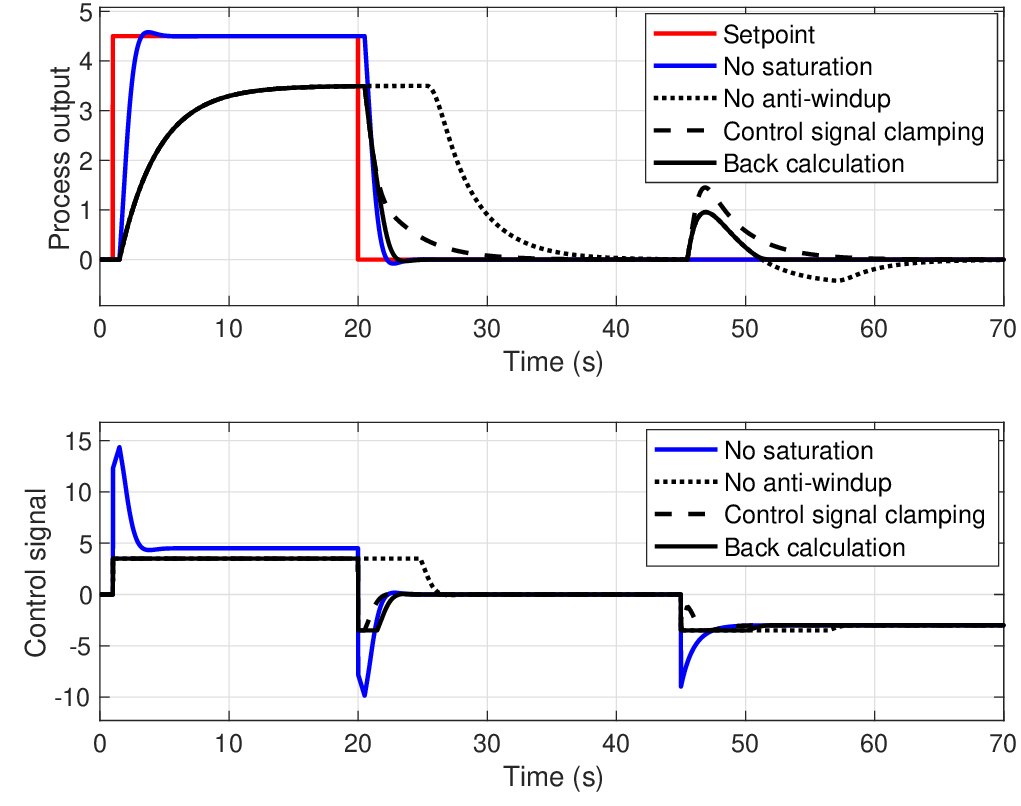}
    \caption{Feed-forward plus saturation in Example~4.}
    \label{fig:example4}
\end{figure}

This example focuses on analysing the anti-windup solutions discussed in \cref{sec:windup} for the control signal saturation problem, and also the combination of the PID controller with a feed-forward compensator to deal with measurable disturbances. The effect of the anti-windup clamping and back calculation solutions implemented in \cref{code:control} will be analysed and compared. Moreover, the capability of the new PID code that includes the feed-forward control signal inside the PID controller will also be explored. Notice that in classical implementations, the contribution of the feed-forward compensator is added to the output of the PID controller outside the controller code \citep{guzman+2024}, and in the proposed code the feed-forward control signal is considered part of the PID controller (see \cref{sec:FF}).\clearpage
For this example, the following process models are used:
\begin{equation}
    P_u(s) = \frac{k_u}{1+sT_u}e^{-sL_u}=\frac{1}{1+3s}e^{-0.5s},
    \label{eq:Pu}
\end{equation}
\begin{equation}
    P_v(s) = \frac{k_v}{1+sT_v}e^{-sL_v}=\frac{2}{1+s}e^{-0.5s},
    \label{eq:Pd}
\end{equation}
\noindent where $P_u$ represents the dynamics that relates the control signal with the process output and $P_v$ represents the dynamics that relates the load disturbance with the process output.
In this case, a PID controller was designed using the Pad\'e approximation for the time delay and using the Lambda method with $\lambda=0.3$, resulting in $K=2.83$, $T_i=3.25$ and $T_d=0.23$. The control signal limits are $u_{\min}=-3.5$ and $u_{\max}=3.5$, and the feed-forward compensator is designed in the classical manner by dividing $P_v$ over $P_u$ with reversed sign:
\begin{equation}
    FF(s) = -\frac{P_v}{P_u}=-\frac{2(1+3s)}{1+s}.
    \label{eq:FF}
\end{equation}
\Cref{fig:example4} shows the simulation results for this example, where the cases without saturation limits (\lstinline{No saturation}) and with saturation limits but without using anti-windup techniques (\lstinline{No anti-windup}) are included for a better comparison. 

The simulation starts with a large setpoint change from $r=0$ to $r=4.5$ at $t=1$. Then, the setpoint value is set again to $r=0$ at $t=20$ and a step-like change is also included for the load disturbance from $v=0$ to $v=1.5$ at $t=45$. Notice that for the no-saturation case, the desired closed-loop response is perfectly achieved for the setpoint changes, and the effect of the disturbance is completely removed.\clearpage
When actuator saturation is taken into account, differences among the evaluated strategies become evident. Between $t = 1$ and $t = 20$, following the first setpoint change, no noticeable differences are observed, since the control signal in all simulated cases exceeds the saturation limit. However, after the second step change at $t = 20$, the impact for the no anti-windup case becomes apparent. In this case, the system performance is significantly degraded, as the controller requires a longer time to exit the saturation region and reach the new setpoint due to the windup effect. The performance is also significantly degraded in the disturbance rejection scenario, where an undesirable overshoot is observed in the response due to the same underlying cause.

\begin{figure}[!t]
    \centering
    \includegraphics[width=0.7\linewidth]{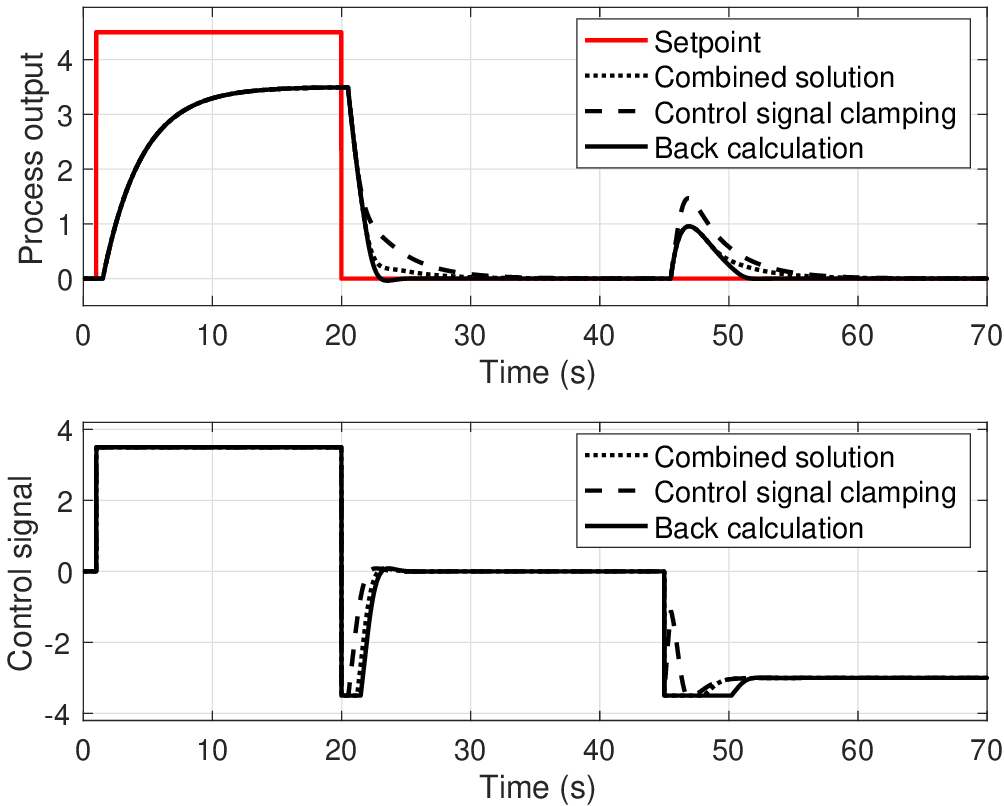}
    \caption{Feed-forward plus saturation in Example~4 comparing anti-windup techniques.}
    \label{fig:example4_b}
\end{figure}

When clamping and back calculation anti-windup techniques are used, the responses are improved in different manners, as observed in the dashed-line and black solid-line curves for clamping and back calculation cases, respectively. As commented in \cref{sec:windup}, the PID code is the same for both techniques only changing the value of $T_t$, with $T_t=\Delta t=0.01$ for clamping and $T_t=T_i=3.25$ for back calculation in this example. When clamping is used, the saturation time is dramatically reduced as observed in the control signal plot. It can be seen that the control signal leaves saturation quickly for both the setpoint tracking and disturbance rejection responses. However, this advantage from a saturation point of view results in a slow response at the process output for setpoint tracking and disturbance rejection, as observed in the process output plot. On the other hand, when anti-windup back calculation is used, the saturation time is only slightly reduced, but the performance on the process output for setpoint tracking and disturbance rejection cases is much closer to the ideal response without saturation limits.
As discussed in \cref{sec:windup}, in this paper a third anti-windup approach is proposed as a combination of clamping and back calculation. \Cref{fig:example4_b} shows the same previous example, where clamping, back calculation, and the combined solution (named \lstinline{Combined solution} in the figure) from \cref{code:anti-windup} are compared. As observed, an intermediate behaviour is obtained in this case, where for setpoint tracking the result is similar to the clamping solution, and for the disturbance rejection problem it is closest to the back calculation solution. The main advantage of this third solution is that the calculated control signal is dynamically adjusted so that the saturation error decays exactly as a first-order process with a time constant $T_t$, whenever the control signal is saturated.
Thus, three different anti-windup options are available to deal with the saturation problem, clamping allowing to leave saturation very fast but slowing down the process output response; back calculation allowing to obtain a trade-off between saturation time and process output performance; and the combination of the two approaches providing an intermediate behaviour according to the trade-off between saturation time and process output performance. Therefore, the user can choose between these three solutions depending on each control problem.
\subsection{Example 5: Noise filtering}\label{sec:example5}
An interesting feature of the implementation presented in this work is that measurement noise filtering is suggested to be combined with the PID controller, as described in \cref{sec:filter}.
\begin{figure}[!ht]
    \centering
    \includegraphics[width=0.7\linewidth]{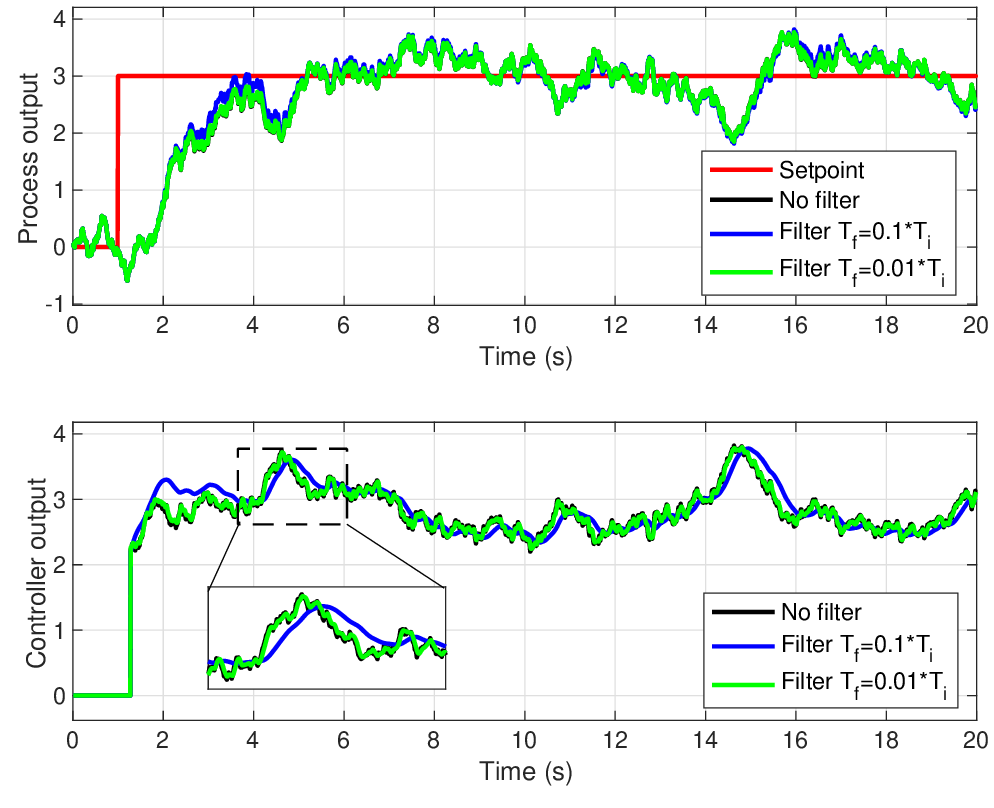}
    \caption{Noise filtering in Example~5.}
    \label{fig:example5}
\end{figure}
This example shows how the process variable can be properly filtered by tuning the $T_f$ parameter as input to the filter function. The process transfer function and PI-controller given by \cref{eq:P} and \cref{eq:C}, respectively, have been used for this simulation, where a white noise signal was added to the process output.
\Cref{fig:example5} shows the simulation result for three cases: without filter, corresponding to $T_f=\Delta t/2$, and with filter for $T_f=0.01T_i$ and $T_f=0.1T_i$, respectively.
\clearpage
The simulation starts with the controller in manual mode with $u=0$. The setpoint is changed from $r=0$ to $r=3$ at $t=1$. At time $t=1.25$ the controller is switched to automatic mode.
As observed in the bottom plot, for the case with $T_f=0.01T_i$, the original control signal is mainly kept and only the very high frequency components of the signal are filtered. For the case with $T_f=0.1T_i$, the original control signal is highly filtered by cutting off high-frequency components. Notice that for all three cases, almost the same process output responses are obtained, which means that the filter is mainly affecting only the control signal.
\subsection{Example 6: Gain scheduling}\label{sec:example6}
\begin{figure}[H]
    \centering
    \includegraphics[width=0.7\linewidth]{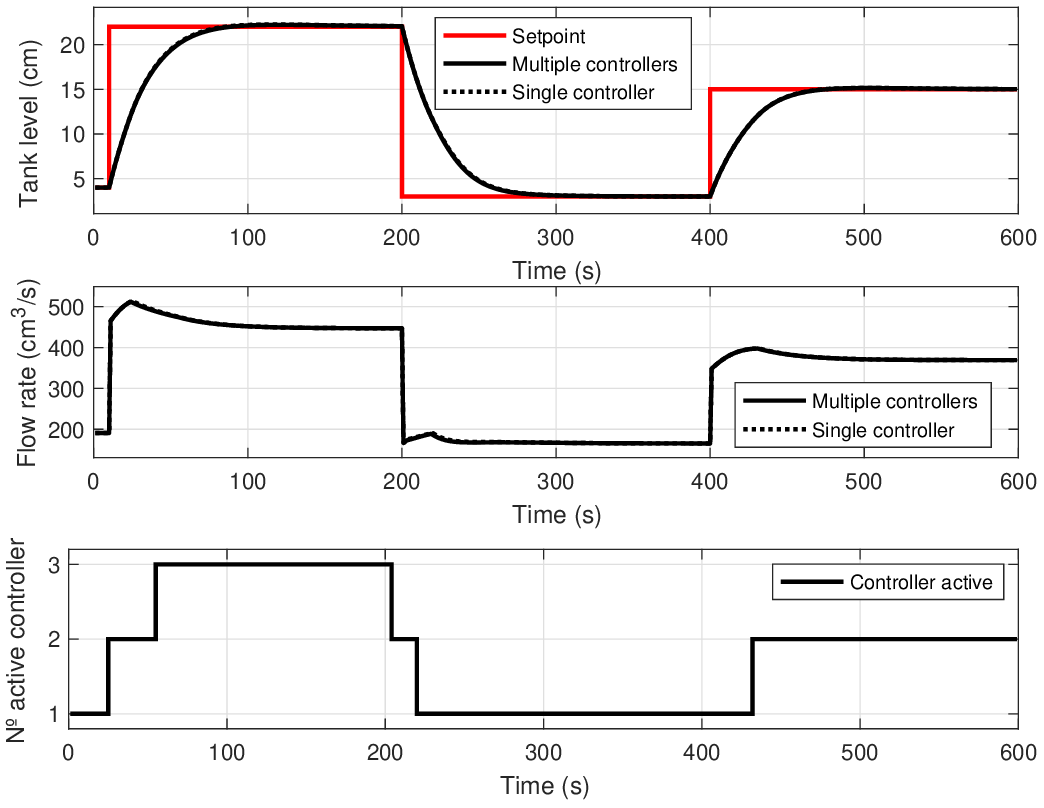}
    \caption{Gain scheduling control results for the tank level control problem in Example~6 for codes \cref{code:example6a} and \cref{code:example6b}.}
    \label{fig:example6}
\end{figure}

This example shows how to use the proposed code for the implementation of gain-scheduling approaches. We consider the classical tank level control problem, whose process dynamics are given by the following differential equation:
\begin{equation}
\frac{dy(t)}{dt}=-\frac{a}{A}\sqrt{2gy(t)}+\frac{u(t)}{A},
\label{eq:tank}
\end{equation}
where $y$ is the tank level, $u$ is the inlet flow, $g=981$~\si{\centi\metre\per\square\second} is the gravitational acceleration, $a=2.15$~\si{\square\centi\metre} is the cross section of the outlet hole, and $A=390$~\si{\square\centi\metre} is the cross section of the tank.
Since the process is nonlinear, it is divided into three different operating zones, where a linear model approximation is derived in each zone. The following three operating points are considered for this study:
\begin{description}
    \item[Zone 1] for $y\in [4,12)$~\si{\centi\metre}
    \item[Zone 2] for $y \in [12,20)$~\si{\centi\metre}
    \item[Zone 3] for $y \in [20,25]$~\si{\centi\metre}
\end{description}
In each zone, the process is modelled as
\begin{equation}
P(s) = \dfrac{k}{1+sT}.
\end{equation}
The model parameters are given in \cref{tab:gainscheduling}.
\begin{table}[t]
\centering
\caption{Process and controller parameters for the gain scheduling Example~6.}
\begin{tabular}{lccc}
\toprule
  &{\bf Zone 1} &{\bf Zone 2} &{\bf Zone 3}\\
\midrule
{\bf Process model parameters} & & &\\
$k$ [\si{\centi\metre}/(\si{\cubic\centi\metre\per\second})] &0.0420 &0.0727 &0.0938\\
$T$ [\si{\second}] &16.36 &28.34 &36.59\\[5pt]
{\bf PI-controller parameters} & & &\\
$K$ [(\si{\cubic\centi\metre\per\second})/\si{\centi\metre}] &14.39 &14.39 &14.39\\
$T_i$ [\si{\second}] &16.36 &28.34 &36.59\\
\bottomrule
\end{tabular}
\label{tab:gainscheduling}
\end{table}
\clearpage
The gain-scheduling control scheme can be implemented in two different ways: by running all the controllers in parallel and switching among them based on the current operating point; or by using a single controller and updating its parameters based on the current operating point. Both solutions have been used in this example, and \cref{code:example6a} and \cref{code:example6b} show the corresponding codes for the multiple controllers and single controller case, respectively. Notice that the initialisation of the variables was omitted and only the control loop is presented. In both cases, the setpoint is increased from $r=4$ to $r=22$~\si{\centi\metre} at $t=10$~\si{\second}, and the process is initialised with a value of $y=4$~\si{\centi\metre} and $u=190.66$~\si{\cubic\centi\metre\per\second}.
When using multiple controllers as presented in \cref{code:example6a}, the key point is to correctly manage the mode status of the controllers, as observed from \lines{ex6amode}{ex6amodeend}. Once the corresponding controller is selected to be active (\lstinline{mode=AUTO}), the rest of the controllers must be in tracking mode, and thus the corresponding variables \lstinline{mode} must be set to \lstinline{TRACK}. Moreover, the tracking signal for all the controllers must be the actual input to the process in order to ensure bumpless transfer when switching among the controllers (see \line{ex6atrack}). In the case of using a single controller as presented in \cref{code:example6b}, the key point is the adaptation of the controller parameters according to the corresponding operating point observed on \lines{ex6bparam}{ex6bparamend}. Notice that in this case, the controller mode is set to \lstinline{AUTO} since the beginning of the simulation.
Both implementations provide identical results. \Cref{fig:example6} shows the simulation results for both cases where the setpoint was changed along the operating points to force the switching among the different controllers. As observed, the same results are obtained in both implementations and the curves in the two upper plots overlap. It can also be seen how the same closed-loop behaviour is obtained despite the changes on the operating range. Moreover, note that no bumps are observed in the control signal at the switching moments. The lower plot shows the number of the active controller, representing the switching state.

\clearpage
\noindent\makebox[\linewidth][c]{
\begin{minipage}[t]{0.48\linewidth}
\begin{lstlisting}[basicstyle=\ttfamily\footnotesize,
  caption={Parallel controllers in Example~6.},
  label={code:example6a}
]
while t < Tsim |\nonum|
    % Simulation of process dynamics |\num|
    y=process_simulation(u,Dt) |\nonum|

    % Controller 1 |\numdecr|     
    u1=pid1.control(r,y,uff,uman,utrack,mode1)|\nonum|
    % Controller 2|\num|
    u2=pid2.control(r,y,uff,uman,utrack,mode2)|\nonum|
    % Controller 3|\num|
    u3=pid3.control(r,y,uff,uman,utrack,mode3)|\nonum|                   

    % Zone 1 |\numdecr|
    if y < 12 |\label{line:ex6amode}|
        mode1="AUTO"
        mode2="TRACK"       
        mode3="TRACK"
        u=u1|\nonum|
    % Zone 2 |\num|
    elseif y >=12 && y < 20
        mode1="TRACK"
        mode2="AUTO"       
        mode3="TRACK"
        u=u2|\nonum|
    % Zone 3 |\num|
    elseif y >=20 
        mode1="TRACK"
        mode2="TRACK"       
        mode3="AUTO"
        u=u3; 
    end|\label{line:ex6amodeend}\nonum|  

    % Tracking signal |\numdecr|
    utrack=u;|\label{line:ex6atrack}\nonum|

    % Set-point change |\numdecr|  
    if t>=200 && t<400
        r=3
    elseif t>=400 && t<600
        r=15
    end |\nonum|       
    |\num|
    t=t+Dt 
end
\end{lstlisting}
\end{minipage}\hspace{0.04\linewidth}
\begin{minipage}[t]{0.48\linewidth}
\begin{lstlisting}[basicstyle=\ttfamily\footnotesize,
  caption={Single controller in Example~6.},
  label={code:example6b}
]
while t < Tsim|\nonum|
    % Simulation of process dynamics|\num|
    y=process_simulation(u,Dt)|\nonum|

    % Controller|\numdecr|    
    u=pid1.control(r,y,uff,uman,utrack,mode)|\nonum|                   

    % Zone 1|\numdecr|
    if y < 12 |\label{line:ex6bparam}|
        pid1.kp=kp1
        pid1.ki=ki1
        pid1.kd=kd1|\nonum| 
    % Zone 2|\num|
    elseif y >=12 && y < 20
        pid1.kp=kp2
        pid1.ki=ki2
        pid1.kd=kd2 |\nonum|
    % Zone 3|\num|
    elseif y >=20 
        pid1.kp=kp3
        pid1.ki=ki3
        pid1.kd=kd3 
    end |\label{line:ex6bparamend}\nonum|              

    % Set-point change|\numdecr|  
    if t>=200 && t<400
        r=3
    elseif t>=400 && t<600
        r=15
    end |\nonum|     
    |\num|
    t=t+Dt
end
\end{lstlisting}
\end{minipage}
}
\subsection{Example 7: Selector or override control}\label{sec:example7}
\begin{figure}[!ht]
    \centering
    \includegraphics[width=0.65\linewidth]{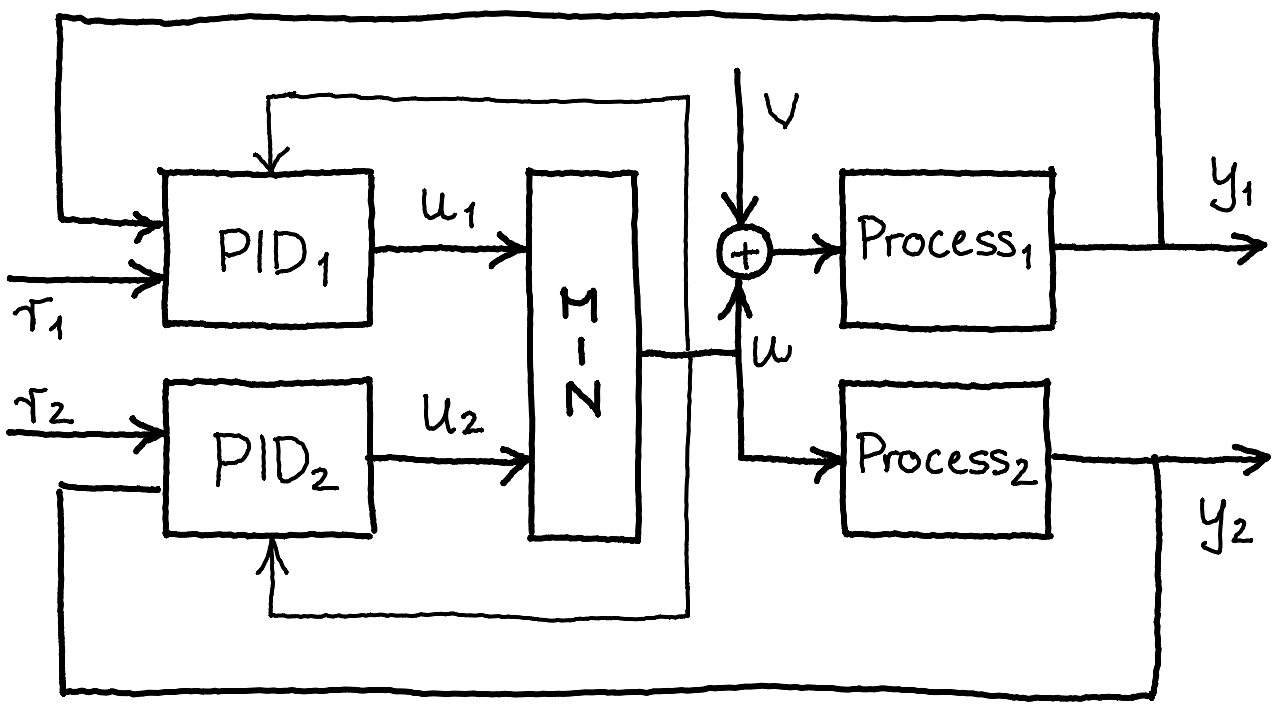}
    \caption{Selector control scheme.}
    \label{fig:selector_scheme}
\end{figure}

\begin{figure}[hb]
    \centering
    \includegraphics[width=0.75\linewidth]{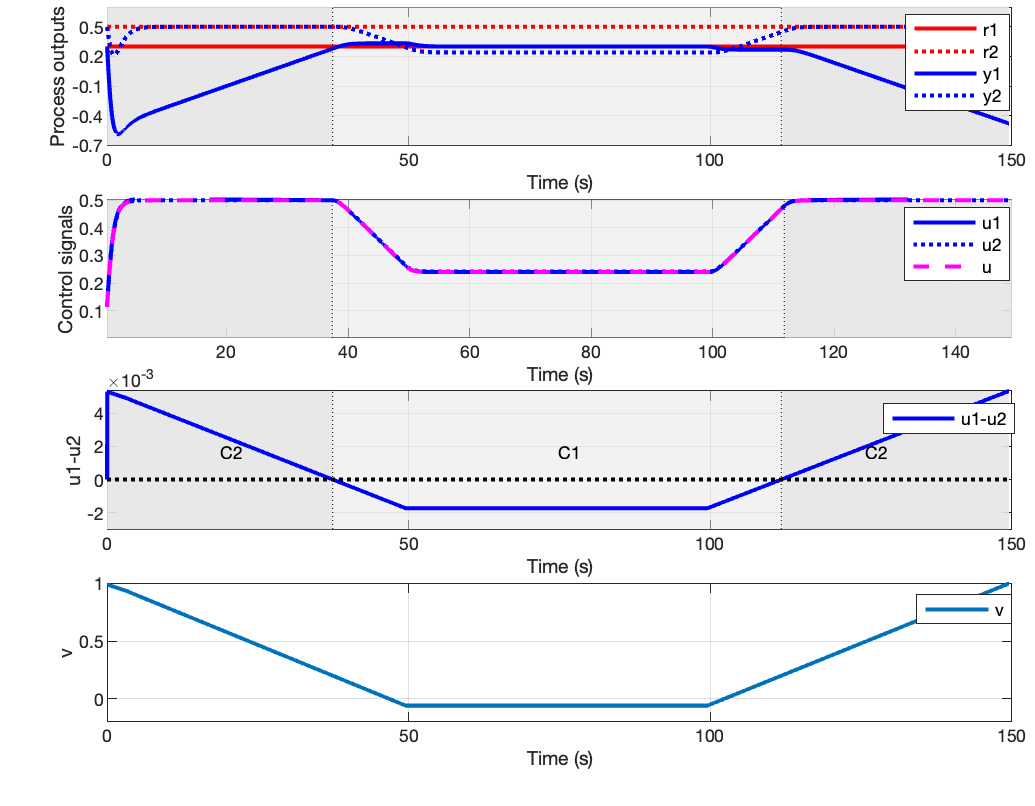}
    \caption{Simulation results for the selector control scheme in Example~7 and for code \cref{code:example7}.}
    \label{fig:example7}
\end{figure}

This example shows how to handle the tracking state when using a selector control approach. The important thing here is to keep the P-part of the controllers as we need their current P-contribution to properly select (for instance, with MIN or MAX selectors) the current control signal.
To analyse this case, we use the example proposed in \cref{fig:selector_scheme}.
The process has one input signal $u$ and two output variables, $y_1$ and $y_2$. The two output variables are input signals to the two controllers $C_1$ and $C_2$. Only one of the controllers is active at a time, in this case the controller with the smallest control signal. The control signal corresponding to the non-active controller is tracking the active control signal.
\clearpage
This structure is common for handling safety constraints. One of the controllers is working in normal operation, but when a certain process variable becomes too small or too large, the other controller takes over, overrides, to ensure that the constraint is not violated.
In this example, $P_1=P_2$ and $C_1=C_2$, where we use the process and controller given by \cref{eq:P} and \cref{eq:C}. The setpoint values are $r_1=0.3$ and $r_2=0.5$.
\Cref{fig:example7} shows the simulation results for this example. A disturbance signal $v$ is subtracted from the input of $P_1$ to generate variations in the operating conditions. The plot at the bottom represents the disturbance signal $v$. As observed from the third plot from the top (representing $u_1-u_2$), the MIN selector works properly. The active control signal is always the smallest one. Moreover,
the controller which is in tracking state calculates the control signal as the current control signal plus its P-part. 
\clearpage
\Cref{code:example7} shows the code in which the handling of the tracking mode signals is observed based on the selector choice. As for the parallel controllers in Example~6, the tracking signal for the two controllers must be the actual input to the process to ensure bumpless transfer when switching among the controllers (see \line{ex7track}).
\begin{lstlisting}[
  caption={Simulation code for selector control approach.},
  label={code:example7}
]
while t < Tsim |\nonum| 
    % Simulation of process 1 dynamics |\num|
    y1=process1_simulation(u-v,Dt) |\nonum| 

    % Simulation of process 2 dynamics |\numdecr|
    y2=process2_simulation(u,Dt) |\nonum| 

    % Controller 1 |\numdecr|
    u1=pid1.control(r1,y1,uff,uman,utrack,mode1) |\nonum|                   

    % Controller 2 |\numdecr|
    u2=pid2.control(r2,y2,uff,uman,utrack,mode2) |\nonum| 

    % Selector |\numdecr|
    if u1 < u2        
        mode1="AUTO"
        mode2="TRACK" 
        u=u1;        
    else        
        mode1="TRACK"
        mode2="AUTO"      
        u=u2        
    end |\nonum|     
    |\num|
    utrack=u |\label{line:ex7track}\nonum|
    |\num|
    t=t+Dt
end



\end{lstlisting}
\vspace*{-1em}
\section{Conclusions}\label{sec:conclusions}
This paper can be used as a practical guide to understand and implement a PID controller in a software environment. The proposed reference implementation is intended to be used as a standard for many applications from the process industry to medical, aeronautical and automotive applications. The contribution lies in the consideration of practical problems that will be encountered when dealing with real-world problems, all of which have been carefully considered and addressed.
The contribution further lies in the discussion of several simulated examples where the different features are tested in closed loop. \rev{The provided insights can be used to implement individual features of our proposed reference implementation, in use cases where it is undesirable to replace the entire {PID} implementation.}
While we expect that small adjustments will have to be made for most applications, we believe that this reference implementation closes the book on PID implementation of the basic form and opens the next when comparing advanced control strategies. In particular, feed-forward control as well as strategies requiring tracking such as selectors or override, gain scheduling and cascade control can be studied in different ways.
We hope that this implementation will also make the use of advanced strategies more popular as these rely on the correct basic implementation, which is not always given in industrial settings.
\clearpage
\thispagestyle{conflictpage}
\bibliographystyle{apacite}
\bibliography{references}
\clearpage
\appendix
\crefalias{section}{appendix}
\section{Pseudo-code language}\label{sec:language}
\Crefrange{code:position}{code:anti-windup} are written in a pseudo-code language similar to MATLAB, a programming language that many control engineers are familiar with, and that the examples, \Crefrange{code:basicloop}{code:example7}, are written in.
Comments are {\color{commentcolor}green}. Keywords are {\color{keywordcolor}blue}. The keywords are:
\begin{itemize}
\item \lstinline{function} to signal that the line contains the signature of a function definition;
\item \lstinline{if}, \lstinline{else}, \lstinline{elseif}, and \lstinline{end} used for logical branching. (Should the target language dictate, \lstinline{else} and \lstinline{elseif} can be replaced by using consecutive \lstinline{if}-\lstinline{end} clauses.) For clarity, we also use the \lstinline{end} keyword to delimit functions.
\end{itemize}
Variables are colour coded, with state variables in {\color{magenta}pink}. These are variables that must retain their values between function calls. They are managed outside the function to allow multiple controller or filter instances, and may be stored using object-oriented or other mechanisms (e.g., as in the MATLAB and Python implementations in our GitHub repository \citep{sundstrom+2026}).
Parameters of the control (\cref{code:control}) and filter (\cref{code:filter}) functions are {\color{amber}orange}. They can be set and updated in various ways, such as being passed as arguments (by value or reference) or retrieved from memory within the function. The specific mechanism is left to the implementer.
To facilitate readability, block indentation is used, but not strictly needed to avoid ambiguity.
Our pseudo-code assumes availability of the standard binary operators \lstinline{==} to check equality and \lstinline{>} to check ordering.
Furthermore, the language is assumed to be equipped with floating point arithmetic operators \lstinline{+}, \lstinline{-}, \lstinline{*}, \lstinline{/}, and the absolute value function \lstinline{abs}.
To facilitate portability, we have intentionally avoided the use of compound data structures such as matrices or lists. Exceptions to this are lists (or tuples) used for arguments and returns of functions.
To define values of the \lstinline{mode} variable we have used macro variables \lstinline{MAN} and \lstinline{TRACK} for readability. A safer option for encoding the mode is proposed in \cref{sec:ControlMode}. Were it not for readability, we could have used the more memory-efficient option of integers (e.g., \lstinline{0} for manual and \lstinline{1} for tracking mode), possibly defined through macros \lstinline{MAN} and \lstinline{TRACK}, language allowing.
We also adhere to a list of semantic rules, to facilitate readability:
\begin{itemize}
\item A leading lowercase \lstinline{d} in a variable name, represents a time derivative. For example \lstinline{dy} is the time derivative of the measurement \lstinline{y}.
\item State variables have variable names with a leading {\color{magenta}\lstinline{x}}. This is to avoid ambiguity between for example the control signal \lstinline{u} and its previous value, stored in \lstinline{xu}.
\item A leading uppercase \lstinline{D} in a variable name represents a difference with respect to the value the variable held in the previous invocation of the function. For example \lstinline{Dy=y-xy} assigns to \lstinline{Dy} the difference between the current measurement \lstinline{y} and its previous value \lstinline{xy}.
\item With the exception of uppercase \lstinline{D} (explained above), variable names are all lowercase.
\end{itemize}
Typing is not explicit, but all numeric values are assumed to automatically be cast. For example \lstinline{0} in \lstinline{ki==0} will be cast into a float \lstinline{0.0} if \lstinline{ki} is of type float.
\clearpage
\section{Signal filter properties}\label{sec:filterapp}
This appendix derives the filter update \cref{eq:dt_filter}, and discusses how the time constant $T_f$ should be interpreted in the contexts of a first-order filter \cref{eq:F1}, and its second-order counterpart \cref{eq:F2}.
If it is assumed that the filter input $y(t)$ is (held) constant between samples, known as zero-order-hold (ZOH), approximation-free closed-form discretisations of \cref{eq:F1} and \cref{eq:F2} can be obtained. This is done using ZOH discretisation, described in for example \citet{astrom+1997}.
Observing that $F_2(s)$ in \cref{eq:F2} is the series connection of two filters of the type $F_1(s)$ in \cref{eq:F1}, it is sufficient to consider discretisation of $F_1(s)$, to then cascade two such discretisations.
We can think of discretisation as a means to approximate the derivative operator $d/dt$ as a function of the shift operator $q$, where $qx(t)=x(t+\Delta t)$. A simple example is the backward difference approximation, which replaces $\dot x(t)$ by $(x(t)-x(t-\Delta t))/\Delta t$. While computationally attractive, this approximation may introduce poor stability and accuracy when $\Delta t$ is not sufficiently small relative to $T_f$. Therefore, we instead employ the Tustin (bilinear) approximation, which can be interpreted directly in the time domain as applying the trapezoidal rule to the time derivative. For a signal $x(t)$ sampled with interval $\Delta t$, the Tustin approximation $\hat{\dot{x}}$ of the derivative $\dot{x}(t)$ can be defined recursively as
\begin{equation}
\hat{\dot{x}}(t)
\approx 2\frac{\Delta x(t)}{\Delta t}-\hat{\dot{x}}(t-\Delta t).
\label{eq:tustin}
\end{equation}
In the transform domain, this corresponds to the bilinear mapping
\begin{equation}
s \approx \frac{2}{\Delta t}\frac{z-1}{z+1}
\label{eq:bilinear}
\end{equation}
between the continuous-time Laplace transform variable $s$ and the discrete-time transform variable $z$.
The continuous-time first-order filter \cref{eq:F1} can be equivalently written in the time domain as the differential equation
\begin{equation}
T_f\dot{y}_f(t)+y_f(t)=y(t),
\label{eq:ct_filter}
\end{equation}
where $y(t)$ is the measured signal and $y_f(t)$ denotes the filtered output.
Substituting the Tustin approximation \cref{eq:tustin} for $\dot{y}_f(t)$ into the continuous-time filter of \cref{eq:ct_filter}, and rearranging terms to isolate $y_f(t)$, yields the discrete-time recursion \cref{eq:dt_filter}, relating the current filtered value to its previous value and the current measurement.
While being computationally favourable, and bumpless when implemented in incremental form, the cascade of two Tustin-approximated filters \cref{eq:tustinF} will have slightly higher noise amplification in the derivative term than the ZOH discretisation of a filtered derivative obtained by applying the time-domain counterpart of $s/(sT_f+1)^2$ to $y(t)$. However, if $n(t)$ is white noise, the difference is negligible for the practically most relevant regime of $\Delta t < 0.1 T_f$. (As $T_f$ decreases to approach $\Delta t$, it can be shown that the noise amplification, in terms of variance, is at most \SI{50}{\percent} better for the ZOH filter.)
It is also worth noting that while $T_f$ is the time constant, in terms of $(1-\exp(-1))\cdot\SI{100}{\percent}\approx\SI{63}{\percent}$ rise time, of $F_1$ in \cref{eq:F1}, the same is not true for the relation between $T_f$ and rise time of $F_2$ in \cref{eq:F2}.\clearpage
If one has tuned a first-order filter \cref{eq:F1} to have a (rise) time constant $T_f'$, and wants to maintain that property for the second order filter \cref{eq:F2}, the following substitution should be made:
\begin{equation}
T_f\approx 2T_f',
\label{eq:Tfapp}
\end{equation}
where the factor of two approximates the solution of
\begin{equation}
\left(1+\frac{T_f'}{T_f}\right)
\exp\!\left(-\frac{T_f'}{T_f}\right)
=
\exp(-1).
\end{equation}
This is of practical importance if employing a tuning method that assumes a first-order filter, when in fact a second-order filter is used, and vice versa.
\clearpage
\vspace*{-5em}
\section{Nomenclature}\label{sec:nomenclature}
This appendix summarises the nomenclature used throughout the paper. \Crefrange{tab:parameter}{tab:variable} list variables from the reference implementation of \crefrange{code:control}{code:anti-windup}, grouped into parameters, states, function arguments, and internal variables.  \Cref{tab:symbol} lists the mathematical symbols used in the theoretical development. For each entry, the ``Listing'' or ``Equation'' column, together with the ``Section'' column, indicates where it appears. An em-dash in the ``Equation'' column means that the symbol is introduced in the running text.
\renewcommand{\arraystretch}{1.2}
\begin{longtable}{p{2cm}p{7cm}ll}
\caption{Parameter variables used in our reference implementation. All parameters are typeset in {\color{amber}orange}.}\label{tab:parameter}\\
\toprule
\textbf{Parameter} & \textbf{Description} & \textbf{Listing} & \textbf{Section} \\
\midrule
\endfirsthead
\toprule
\textbf{Parameter} & \textbf{Description} & \textbf{Listing} & \textbf{Section} \\
\midrule
\endhead
\lstinline{b} & Setpoint weight $b$ for proportional term & \ref{code:control} & \ref{sec:setpoint} \\
\lstinline{c} & Setpoint weight $c$ for derivative term & \ref{code:control} & \ref{sec:setpoint} \\
\lstinline{Dt} & Execution interval $\Delta t$ & \ref{code:control} & \ref{sec:jitter} \\
\lstinline{dumax} & Upper control signal rate limit $\dot{u}_{\max}$ & \ref{code:control} & \ref{sec:ulims} \\
\lstinline{dumin} & Lower control signal rate limit $\dot{u}_{\min}$ & \ref{code:control} & \ref{sec:ulims} \\
\lstinline{kd} & Derivative gain $k_d=k_p\ T_d$ & \ref{code:control} & \ref{sec:positional} \\
\lstinline{ki} & Integral gain $k_i=k_p/T_i$ & \ref{code:control} & \ref{sec:positional} \\
\lstinline{kp} & Proportional gain $k_p=K$ & \ref{code:control} & \ref{sec:positional} \\
\lstinline{Tf} & Filter time constant & \ref{code:filter} & \ref{sec:filter} \\
\lstinline{Tt} & Back calculation anti-windup time constant & \ref{code:control} & \ref{sec:windup} \\
\lstinline{u0} & Control signal bias term $u_0$ & \ref{code:control} & \ref{sec:PandPD} \\
\lstinline{umax} & Upper control signal saturation limit $u_{\max}$ & \ref{code:control} & \ref{sec:ulims} \\
\lstinline{umin} & Lower control signal saturation limit $u_{\min}$ & \ref{code:control} & \ref{sec:ulims} \\
\bottomrule
\end{longtable}
\begin{longtable}{p{2cm}p{7cm}ll}
\caption{State variables used in our reference implementation.}\label{tab:state}\\
\toprule
\textbf{State} & \textbf{Description} & \textbf{Listing} & \textbf{Section} \\
\midrule
\endfirsthead
\toprule
\textbf{State} & \textbf{Description} & \textbf{Listing} & \textbf{Section} \\
\midrule
\endhead
\lstinline{xdr} & Time derivative of setpoint & \ref{code:control} & \ref{sec:code-derivatives} \\
\lstinline{xdy} & Time derivative of process variable & \ref{code:control} & \ref{sec:code-derivatives} \\
\lstinline{xf1} & Filter state 1 & \ref{code:filter} & \ref{sec:code-filter} \\
\lstinline{xf2} & Filter state 2 & \ref{code:filter} & \ref{sec:code-filter} \\
\lstinline{xr} & Setpoint & \ref{code:control} & \ref{sec:code-stateupdate} \\
\lstinline{xu} & Unsaturated control signal & \ref{code:control} & \ref{sec:code-integral} \\
\lstinline{xuff} & Feed-forward control signal & \ref{code:control} & \ref{sec:FF} \\
\lstinline{xus} & Saturated control signal & \ref{code:control} & \ref{sec:code-saturation} \\
\lstinline{xy} & Process variable & \ref{code:control} & \ref{sec:code-stateupdate} \\
\bottomrule
\end{longtable}
\begin{longtable}{p{2cm}p{7cm}ll}
\caption{Function argument variables passed to the control function.}\label{tab:argument}\\
\toprule
\textbf{Argument} & \textbf{Description} & \textbf{Listing} & \textbf{Section} \\
\midrule
\endfirsthead
\toprule
\textbf{Argument} & \textbf{Description} & \textbf{Listing} & \textbf{Section} \\
\midrule
\endhead
\lstinline{mode} & Controller operating mode & \ref{code:control} & \ref{sec:ControlMode} \\
\lstinline{r} & Setpoint & \ref{code:control} & \ref{sec:CombinedPID} \\
\lstinline{uff} & Feed-forward control signal & \ref{code:control} & \ref{sec:FF} \\
\lstinline{uman} & Manual control signal & \ref{code:control} & \ref{sec:code-man} \\
\lstinline{utrack} & Tracking control signal & \ref{code:control} & \ref{sec:tracking} \\
\lstinline{y} & Process variable & \ref{code:control} & \ref{sec:CombinedPID} \\
\bottomrule
\end{longtable}
\begin{longtable}{p{2cm}p{7cm}ll}
\caption{Other variables used in \cref{code:control,code:filter}.}\label{tab:variable}\\
\toprule
\textbf{Variable} & \textbf{Description} & \textbf{Listing} & \textbf{Section} \\
\midrule
\endfirsthead
\toprule
\textbf{Variable} & \textbf{Description} & \textbf{Listing} & \textbf{Section} \\
\midrule
\endhead
\lstinline{a} & Filter coefficient & \ref{code:filter} & \ref{sec:code-filter} \\
\lstinline{Ddr} & Setpoint derivative increment & \ref{code:control} & \ref{sec:code-integral} \\
\lstinline{Ddy} & Process variable derivative increment & \ref{code:control} & \ref{sec:code-integral} \\
\lstinline{Dr} & Setpoint increment & \ref{code:control} & \ref{sec:code-integral} \\
\lstinline{Duff} & Feed-forward increment & \ref{code:control} & \ref{sec:code-integral} \\
\lstinline{Dud} & Derivative increment & \ref{code:control} & \ref{sec:code-integral} \\
\lstinline{Dui} & Integral increment & \ref{code:control} & \ref{sec:code-integral} \\
\lstinline{Dup} & Proportional increment & \ref{code:control} & \ref{sec:code-integral} \\
\lstinline{Dy} & Process variable increment & \ref{code:control} & \ref{sec:code-integral} \\
\lstinline{dr} & Setpoint derivative & \ref{code:control} & \ref{sec:code-derivatives} \\
\lstinline{dy} & Process variable derivative & \ref{code:control} & \ref{sec:code-derivatives} \\
\lstinline{u} & Control signal & \ref{code:control} & \ref{sec:code-integral} \\
\lstinline{us} & Saturated control signal & \ref{code:control} & \ref{sec:code-saturation} \\
\lstinline{usmax} & Upper saturation bound & \ref{code:control} & \ref{sec:code-ratelim} \\
\lstinline{usmin} & Lower saturation bound & \ref{code:control} & \ref{sec:code-ratelim} \\
\bottomrule
\end{longtable}
\clearpage
\begin{longtable}{p{2cm}p{7cm}ll}
\caption{Mathematical symbols used throughout the paper.}\label{tab:symbol}\\
\toprule
\textbf{Symbol} & \textbf{Description} & \textbf{Equation} & \textbf{Section} \\
\midrule
\endfirsthead
\toprule
\textbf{Symbol} & \textbf{Description} & \textbf{Equation} & \textbf{Section} \\
\midrule
\endhead
$a$ & Filter coefficient in discrete-time low-pass filter & \ref{eq:a} & \ref{sec:filter} \\
$b$ & Setpoint weight for proportional action & \ref{eq:pidTimeDomain2} & \ref{sec:setpoint} \\
$c$ & Setpoint weight for derivative action & \ref{eq:pidTimeDomain2} & \ref{sec:setpoint} \\
$C(s)$ & Controller transfer function & \ref{eq:C} & \ref{sec:basicexamples} \\
$\Delta e(t)$ & Increment of control error & \ref{eq:approximations} & \ref{sec:sampling} \\
$\Delta e(t-\Delta t)$ & Previous error increment & \ref{eq:incformLong} & \ref{sec:incremental} \\
$\Delta n$ & Increment of measurement noise & --- & \ref{sec:filter} \\
$\Delta r(t)$ & Increment of setpoint & \ref{eq:incformLongSP} & \ref{sec:setpoint} \\
$\Delta t$ & Execution interval / sampling interval & \ref{eq:approximations} & \ref{sec:sampling} \\
$\Delta u(t)$ & Control signal increment & \ref{eq:incrementLaw} & \ref{sec:incremental} \\
$\Delta u_d(t)$ & Derivative contribution increment & \ref{eq:incformLong} & \ref{sec:incremental} \\
$\Delta u_i(t)$ & Integral contribution increment & \ref{eq:incformLong} & \ref{sec:incremental} \\
$\Delta u_p(t)$ & Proportional contribution increment & \ref{eq:incformLong} & \ref{sec:incremental} \\
$\Delta u_{ff}(t)$ & Feed-forward control increment & \ref{eq:Duff} & \ref{sec:FF} \\
$\Delta y(t)$ & Increment of process variable & \ref{eq:incformLongSP} & \ref{sec:setpoint} \\
$\dot{u}(t)$ & Time derivative of control signal & \ref{eq:ratelim} & \ref{sec:ulims} \\
$\dot{u}_{\max}$ & Upper control signal rate limit & \ref{eq:ratelim} & \ref{sec:ulims} \\
$\dot{u}_{\min}$ & Lower control signal rate limit & \ref{eq:ratelim} & \ref{sec:ulims} \\
$\dot{x}(t)$ & Time derivative of signal $x(t)$ & \ref{eq:tustin} & \ref{sec:filterapp} \\
$\dot{y}_f(t)$ & Time derivative of filtered process variable & \ref{eq:ct_filter} & \ref{sec:filterapp} \\
$e(t)$ & Control error & \ref{eq:pidTimeDomain} & \ref{sec:digimp} \\
$F_1(s)$ & First-order filter transfer function & \ref{eq:F1} & \ref{sec:filter} \\
$F_2(s)$ & Second-order filter transfer function & \ref{eq:F2} & \ref{sec:filter} \\
$FF(s)$ & Feed-forward compensator transfer function & \ref{eq:FF} & \ref{sec:example4} \\
$g$ & Gravitational acceleration & \ref{eq:tank} & \ref{sec:example6} \\
$j$ & Summation index & \ref{eq:approximations} & \ref{sec:sampling} \\
$k$ & Process gain & \ref{eq:P} & \ref{sec:basicexamples} \\
$k_d$ & Derivative gain & \ref{eq:posForm1} & \ref{sec:positional} \\
$k_i$ & Integral gain & \ref{eq:posForm1} & \ref{sec:positional} \\
$k_p$ & Proportional gain & \ref{eq:posForm1} & \ref{sec:positional} \\
$k_u$ & Gain of process from control signal to output & \ref{eq:Pu} & \ref{sec:example4} \\
$k_v$ & Gain of process from disturbance to output & \ref{eq:Pd} & \ref{sec:example4} \\
$K$ & Controller gain & \ref{eq:pidTimeDomain} & \ref{sec:digimp} \\
$L$ & Process dead time & \ref{eq:P} & \ref{sec:basicexamples} \\
$L_u$ & Dead time of process from control signal to output & \ref{eq:Pu} & \ref{sec:example4} \\
$L_v$ & Dead time of process from disturbance to output & \ref{eq:Pd} & \ref{sec:example4} \\
$n$ & Discrete-time index & --- & \ref{sec:sampling} \\
$n(t)$ & Measurement noise signal & --- & \ref{sec:filter} \\
$P(s)$ & Process transfer function & \ref{eq:P} & \ref{sec:basicexamples} \\
$P_u(s)$ & Process transfer function from control signal to output & \ref{eq:Pu} & \ref{sec:example4} \\
$P_v(s)$ & Process transfer function from disturbance to output & \ref{eq:Pd} & \ref{sec:example4} \\
$q$ & Shift operator & --- & \ref{sec:filterapp} \\
$r(t)$ & Setpoint signal & \ref{eq:pidTimeDomain} & \ref{sec:digimp} \\
$r_1$ & Setpoint for controller 1 & --- & \ref{sec:example7} \\
$r_2$ & Setpoint for controller 2 & --- & \ref{sec:example7} \\
$s$ & Laplace transform variable & \ref{eq:F1} & \ref{sec:filter} \\
$T$ & Process time constant & \ref{eq:P} & \ref{sec:basicexamples} \\
$T_d$ & Derivative time & \ref{eq:pidTimeDomain} & \ref{sec:digimp} \\
$T_f$ & Filter time constant & \ref{eq:F1} & \ref{sec:filter} \\
$T_f'$ & Equivalent first-order filter time constant & \ref{eq:Tfapp} & \ref{sec:filterapp} \\
$T_i$ & Integral time & \ref{eq:pidTimeDomain} & \ref{sec:digimp} \\
$T_t$ & Anti-windup tracking time constant & --- & \ref{sec:windup} \\
$T_u$ & Time constant of process from control signal to output & \ref{eq:Pu} & \ref{sec:example4} \\
$T_v$ & Time constant of process from disturbance to output & \ref{eq:Pd} & \ref{sec:example4} \\
$t$ & Continuous time & \ref{eq:pidTimeDomain} & \ref{sec:digimp} \\
$u(t)$ & Control signal & \ref{eq:pidTimeDomain} & \ref{sec:digimp} \\
$u_d(t)$ & Derivative control signal term & \ref{eq:posForm2} & \ref{sec:positional} \\
$u_i(t)$ & Integral control signal term & \ref{eq:posForm2} & \ref{sec:positional} \\
$u_p(t)$ & Proportional control signal term & \ref{eq:posForm2} & \ref{sec:positional} \\
$u_0$ & Control signal bias term & \ref{eq:posFormBias} & \ref{sec:PandPD} \\
$u_{ff}(t)$ & Feed-forward control signal & \ref{eq:Duff} & \ref{sec:FF} \\
$u_{\text{man}}(t)$ & Manual control signal & --- & \ref{sec:code-man} \\
$u_{\text{track}}(t)$ & Tracking control signal & \ref{eq:track} & \ref{sec:tracking} \\
$u_{\max}$ & Upper control signal limit & \ref{eq:ulimits} & \ref{sec:ulims} \\
$u_{\min}$ & Lower control signal limit & \ref{eq:ulimits} & \ref{sec:ulims} \\
$u_s(t)$ & Saturated control signal & \ref{eq:uIncSat} & \ref{sec:ulims} \\
$u_{s,\max}$ & Effective upper saturation limit & \ref{eq:usmax} & \ref{sec:ulims} \\
$u_{s,\min}$ & Effective lower saturation limit & \ref{eq:usmin} & \ref{sec:ulims} \\
$v(t)$ & Load disturbance & --- & \ref{sec:FF} \\
$x(t)$ & Generic signal used in filter derivation & \ref{eq:tustin} & \ref{sec:filterapp} \\
$y(t)$ & Process variable (measurement) & \ref{eq:pidTimeDomain} & \ref{sec:digimp} \\
$y_f(t)$ & Filtered process variable & \ref{eq:dt_filter} & \ref{sec:filter} \\
$y_1$ & First process output & --- & \ref{sec:example7} \\
$y_2$ & Second process output & --- & \ref{sec:example7} \\
$z$ & Discrete-time transform variable & \ref{eq:bilinear} & \ref{sec:filterapp} \\
$\hat{\dot{x}}(t)$ & Tustin approximation of $\dot{x}(t)$ & \ref{eq:tustin} & \ref{sec:filterapp} \\
$\lambda$ & Lambda tuning parameter & \ref{eq:C} & \ref{sec:basicexamples} \\
$\omega_c$ & Filter cutoff frequency & \ref{eq:F1} & \ref{sec:filter} \\
\bottomrule
\end{longtable}
\end{document}